\documentclass[11pt]{article}
\usepackage[letterpaper,margin=1.0in]{geometry}
\usepackage{algorithm,algpseudocode,booktabs,comment}
\usepackage{mathtools}

\usepackage{amsmath,amsthm}
\usepackage[table]{xcolor}
\usepackage{array}
\usepackage{booktabs}
\usepackage{diagbox}
\usepackage{tikz}
\usetikzlibrary{arrows.meta,positioning}

\usepackage{amssymb}
\newtheorem{definition}{Definition}
\newcommand{\Description}[1]{}  

\definecolor{cSAME}{RGB}{200,230,201}
\definecolor{cDIFF}{RGB}{197,217,241}
\definecolor{cSTUCK}{RGB}{245,198,203}
\definecolor{cSPEC}{RGB}{255,224,178}

\newcolumntype{Y}{>{\centering\arraybackslash}p{0.95cm}}
\usepackage{graphicx}
\usepackage{makecell}
\usepackage{tabularx}
\usepackage{thmtools,thm-restate}
\DeclareMathOperator*{\argmin}{argmin}

\usetikzlibrary{arrows.meta,shapes.geometric,positioning,fit,backgrounds}

\definecolor{cState}{RGB}{123,31,162}   
\definecolor{cCond}{RGB}{245,124,0}     
\definecolor{cAct}{RGB}{56,142,60}      
\definecolor{cA}{RGB}{56,142,60}        
\definecolor{cB}{RGB}{25,118,210}       
\definecolor{cX}{RGB}{211,47,47}        
\definecolor{cS}{RGB}{245,124,0}        

\usepackage{textcomp}
\usepackage{xcolor}
\usepackage{enumitem}
\usepackage{float}
\usepackage[font=scriptsize]{caption}
\usepackage{wrapfig}
\usepackage[labelformat=simple]{subcaption}

\usepackage{pgf,tikz,pgfplots}
\pgfplotsset{compat=1.15}
\usepackage{mathrsfs}
\usepackage{hyperref}
\usepackage{multicol}
\usepackage{multirow}

\usepackage{xcolor}
\usepackage[linesnumbered,ruled,vlined,algo2e]{algorithm2e}
\usepackage[export]{adjustbox}

\SetCommentSty{mycommfont}

\newtheorem{lemma}{Lemma}
\newtheorem{theorem}{Theorem}
\newtheorem{corollary}{Corollary}
\newtheorem{observation}{Observation}

\usepackage{balance} 

\title{
Beyond 4: Improved Competitive Traveling Repairperson} 

\author{
Bibek Maharjan\\
\small Kent State University\\
\small\texttt{bmaharj1@kent.edu}
\and
Gokarna Sharma\\
\small Kent State University\\
\small\texttt{gsharma2@kent.edu}
}
\date{}

\begin{document}

\maketitle

\begin{abstract}
With $k\geq 1$ repairpersons starting initially from a depot, we present a 3.877-competitive deterministic algorithm for the fundamental online traveling repairperson problem on an arbitrary metric. For the special class of zero-residual instances, we improve the ratio to $3+\epsilon$ for every $\epsilon>0$.  We also extend these results to some variants of the related so-called online dial-a-ride problem. The state-of-the-art is the 4-competitive deterministic algorithm for both problems.  
\end{abstract}



\section{Introduction}
\label{sec:Introduction}

The online Traveling Repairperson Problem (TRP) 
is a fundamental problem in online routing  \cite{BienkowskiKL21-ICALP,bonifaci2009-TCS,feuerstein2001line,Hwang}.  In its standard formulation, requests are released over time at locations in a metric space, and a single (at most) unit-speed repairperson (aka server or vehicle) starting from a designated {\em depot} must visit every request. Each request $r$ has a release time $a_r\geq 0$, a location $x_r$, and a weight $w_r>0$. 
Its completion time $C_r$ is the first time at which the  server visits $x_r$ after the request has been released.
The objective is to minimize the {\em total weighted completion time}, $\sum_r w_r C_r.$ Unlike the online Traveling Salesperson Problem (TSP), which seeks to minimize the {\em maximum weighted completion time}, $\max_r w_rC_r$, TRP emphasizes serving requests quickly on average. Therefore, TRP is particularly relevant to applications in which responsiveness to individual requests is important. 
The {\em competitive ratio} compares the cost $\mathrm{ALG}$ of an online algorithm against the cost $\mathrm{OPT}$ of an optimal offline algorithm that knows the entire request sequence in advance \cite{Borodin98}.


The central challenge in online TRP 
is that the server must balance the service of currently known requests against the possibility that new requests will arrive while it is traveling. Committing to a route computed from the currently available requests may cause substantial delays for future arrivals. Conversely, repeatedly changing the route can prevent the server from completing requests efficiently. Designing algorithms that balance these competing considerations is therefore a central issue in the competitive analysis of online TRP 
\cite{BienkowskiKL21-ICALP,bonifaci2009-TCS,feuerstein2001line,Hwang}.

When $k\geq 1$ servers are available, 
we can denote the problem by $k$-TRP. 
For $k$-TRP, 
the state-of-the-art deterministic lower bound is 2  for $k\geq 2$ \cite{bonifaci2009-TCS}. For $1$-TRP, the state-of-the-art deterministic lower bound is $1+\sqrt{2}\approx 2.414$ \cite{feuerstein2001line}. 
In this paper, with $k\geq 1$, we consider designing deterministic algorithms for $k$-TRP on an arbitrary metric space. The state-of-the-art is the deterministic 4-competitive algorithm  due to Bienowski, Kraska, and Liu 
\cite{BienkowskiKL21-ICALP}. 

%
%
%
%

\vspace{2mm}
\noindent{\bf Contribution.}
We establish the following main result. 

\begin{theorem}
\label{theorem:main}
For any $k\geq 1$, there is a $3.877$\footnote{We use throughout the paper the rounding of the non-integer competitive ratios to three decimal places.}-competitive deterministic online algorithm for $k$-TRP  on an arbitrary metric. The ratio can be improved to $3 +\epsilon$, for every $\epsilon>0$, 
for the special class of zero-residual instances.
\end{theorem}
Theorem \ref{theorem:main} is interesting as it narrows the gap between the deterministic upper and lower bounds \cite{BienkowskiKL21-ICALP,bonifaci2009-TCS,feuerstein2001line} and 
opens the door for future research on techniques to obtain better bounds. 


\vspace{2mm}
{\bf \em Extension to Dial-A-Ride Problem.}
We extend Theorem \ref{theorem:main} to some variants of the so-called Dial-A-Ride Problem (DARP), a  natural and well-studied extension of $k$-TRP,  in which each request has a source and a destination and the goal is to transport a passenger between these two points. 
Therefore, the location $x_r$ for a TRP request $r$ needs to be denoted with two locations: pickup location $s_r$ and delivery location $d_r$. Request $r$ is complete only when its passenger reaches  $d_r$.
Each server may have a {\em finite} capacity limiting how many passengers it may carry at any time (i.e., {\em capacitated}); this capacity may be also infinite (i.e., {\em uncapacitated}). Additionally, in a {\em preemptive} variant, passengers can be unloaded at some points of the metric space (different than their destination), whereas in a {\em non-preemptive} variant, such unloading is not allowed. In an uncapacitated $k$-DARP, preemption is immaterial. 
For any variant of online $k$-DARP, the state-of-the-art upper and lower bounds are the same as online $k$-TRP discussed above, with the only exception that, for capacitated $1$-DARP with server capacity 1, the lower bound increases to 3 \cite{feuerstein2001line}. In this paper, we extend the 3.877 ratio 
to the uncapacitated $k$-DARP. We extend the $3+\epsilon$ ratio to any non-preemptive variant for the special class of zero-residual instances.
Interestingly, this bound almost matches the $3$-competitive lower bound \cite{feuerstein2001line} for $1$-DARP with server capacity 1, the first such result. 
We focus from now on on $k$-TRP and  discuss all DARP aspects in Section 
\ref{sec:k-DARP}.

\vspace{2mm}
\noindent{\bf Overview of the Techniques.}
We call our framework the {\em interruptible capped-route} framework which we discuss here in the context of $1$-TRP for simplicity. 
Our framework builds on the capped-route framework developed by Bienkowski, Kraska, and Liu \cite{BienkowskiKL21-ICALP}. 
As in \cite{BienkowskiKL21-ICALP}, our framework works in phases of geometrically increasing lengths. In each phase, 
the main idea is to compute an {\em optimal capped route} (with one end at the depot) over the requests revealed so far, minimizing the sum of weighted completion times with an additional penalty for unserved requests.   
In contrast to \cite{BienkowskiKL21-ICALP} which, in each phase, executes the entire capped route,
our framework does not necessarily execute the entire route. Instead, it monitors the requests arriving during the execution of the current route in that phase and decides whether continuing the current route is worthwhile. 
Our framework introduces two complementary interruption mechanisms.

\begin{itemize}
    \item An {\em emergency interruption rule} reacts to sufficiently large amounts of newly released request weight. When the accumulated weight of new requests is large compared with the weight of the outstanding requests in the current route, the server interrupts the current route and returns to the depot. 

\item A {\em scheduled interruption rule} examines the balance between the requests outside the current route and the not-yet-traversed tail of that route. If the weight outside the route is sufficiently large relative to the tail weight, the server cuts the route early. Otherwise, it completes the current route before returning to the depot.
\end{itemize}

The analysis relies on two complementary accounting mechanisms: virtual cost and physical cost.  
The {\em virtual cost} captures an accounting value assigned to each request when it is first-registered. 
Its purpose is to connect the physical cost $\mathrm{ALG}$ of the online algorithm 
to the optimal offiline cost $\mathrm{OPT}$. 
We say a request is {\em registered} in a phase if it is included in the route computed at that phase.   
We say a request becomes {\em first-registered} in a phase when its location has been reached by the server in that phase while traversing the current route.  Therefore, a request registered in a phase may become first-registered either in that phase or in a later phase. When a request becomes first-registered in a phase, we denote by {\em virtual registration cost} as the accounting cost we assign to that request.
Therefore, for the virtual cost accounting, we define the {\em capped objective}
$\mathcal{L}(h)=\min_{P}\sum_r w_r\min\{c_{Pr},h\}$,
where $c_{Pr}$ denotes the {\em virtual service time} of request $r$ under capped route $P$ which is the total time to reach the location $x_r$ of $r$ following route $P$ starting from the depot. 
Differences between the optimal capped-route costs at consecutive phase horizons (start times), 
$\Delta_j=L(h_j)-L(h_{j-1}),
$
provide a budget for the loss caused by interruptions (emergency or scheduled). In particular, let $e_j$ denote the capped-route cost benefit sacrificed in phase $j$ due to interruptions (aka {\em residual cost}), we establish a phase-wise bound of the form
$e_j\le \zeta\Delta_j$ with $\zeta$ being some positive constant.
Since the phase horizons are geometrically increasing, the capped-route cost increments $\Delta_j$ {\em telescope}. This yields a global bound on the total interruption loss $e_j$ and, consequently, on the total virtual registration cost, denoted by $B$, of the requests.

The {\em physical cost} is the actual weighted completion time $\mathrm{ALG}$ incurred by the server, 
i.e., $\mathrm{ALG}=\sum_rw_r C_r$.
The physical cost analysis uses a different charging mechanism so that it can be related to the virtual registration cost. Requests registered late in a completed phase receive sufficient unused virtual-cost allowance to compensate for the excess physical cost of requests that may be registered in the following phase. This allows us to bound $\mathrm{ALG}$ in terms of the virtual registration cost. In particular, we establish the bound of the form $B\leq \mathrm{OPT}+\sum_j e_j$, where $B$ is the total virtual registration cost we assign to the requests. 
Notice that the virtual registration cost is the accounting tool; the physical cost is what the algorithm actually pays. 

We indeed show that the actual physical cost $\mathrm{ALG}$ can be charged to the total virtual registration cost $B$ within an overall multiplier of $K$. 
We also show that $B$ 
can be charged to the optimal offline cost $\mathrm{OPT}$ with an overall multiplier of $(1+\gamma)$, and hence the competitive ratio becomes $\frac{\mathrm{ALG}}{\mathrm{OPT}}  \approx K(1+\gamma)$. In fact, we establish the following 
$\frac{\mathrm{ALG}}{\mathrm{OPT}}\leq \frac{2(4\alpha^3-2\alpha^2-\alpha+1)}{\alpha(2\alpha-1)}$ with $\alpha\in (1/2,1)$ using  rational parameter choices for $K$ and $\gamma$
together with the corresponding emergency interruption threshold $M=\frac{\theta}{(1-\alpha)}$ and scheduled interruption threshold $\theta$ to make the emergency and scheduled interruption bounds equal.  
Setting $\alpha^*\approx0.892$ gives 
$\min_{\alpha} \frac{\mathrm{ALG}}{\mathrm{OPT}}\approx 3.877.$ Interestingly, setting $\alpha=1$ (which means an uninterruptible framework like \cite{BienkowskiKL21-ICALP}), our framework gives $\frac{\mathrm{ALG}}{\mathrm{OPT}}\leq \frac{2(4\alpha^3-2\alpha^2-\alpha+1)}{\alpha(2\alpha-1)}=4$, matching the competitive ratio of \cite{BienkowskiKL21-ICALP}.    
With $\alpha=\alpha^*$, the next phase horizon becomes a  multiplicative increase of $1+2\alpha \approx 2.784$ w.r.t. the current phase horizon, following an interruption (emergency or scheduled) in the current phase. If no interruption in the current phase, as in \cite{BienkowskiKL21-ICALP}, the next phase horizon  becomes a multiplicative increase of $3$. 

{\bf\em  Special Class: Zero-Residual Instances.}
We call the instances {\em zero-residual} class if the capped-route cost loss (residual cost) due to  interruptions (emergency or scheduled) becomes zero in our framework, i.e.,    
$e_j=0$ for every phase $j$. A special case of this class are strictly {\em interruption-free} instances in which the emergency and scheduled interruptions are never triggered.  
This means that whenever the capped route $P$ in a  phase selects a request, that request is registered before the phase ends. We show that (i) the additive $\sum_je_j$ term disappears from $B$ and (ii) the actual physical cost $\mathrm{ALG}$ can still be charged to $B$ with an overall multiplier of $K$. Therefore, the ratio becomes $\frac{\mathrm{ALG}}{\mathrm{OPT}}\leq (2+2\alpha)$ with $\alpha\in (1/2,1).$ Setting $\alpha=1/2+\epsilon, \epsilon>0,$ gives $\frac{\mathrm{ALG}}{\mathrm{OPT}}= 3+\epsilon.$    

Finally, with appropriate modifications, we show that our framework extends  to $k$-TRP 
guaranteeing the same deterministic competitive ratios.

\vspace{2mm}
\noindent{\bf Related Work.}
The literature on TSP 
is dense with both deterministic and randomized competitive ratio guarantees. We only discuss deterministic guarantees for $k$-TSP 
as randomized guarantees 
are not the focus of this paper. We sometime simply write TRP 
to denote 1-TRP.

Online TRP was first studied by Feuerstein and Stougie \cite{feuerstein2001line}. They considered the problem on the line metric and obtained a deterministic $9$-competitive algorithm by adapting ideas from the {\em cow-path} problem (aka the linear search problem). They also established a deterministic lower bound of $1+\sqrt{2}=2.414$ for online TRP. 

Krumke {\it et al.} \cite{KrumkePPS03-TCS} subsequently developed a phase-based approach for online TRP on an arbitrary metric. Their deterministic algorithm achieves a competitive ratio of
$(1+\sqrt{2})^2=3+2\sqrt{2}<5.829.$
In their phase-based framework (which was extended by $\cite{BienkowskiKL21-ICALP}$), the server repeatedly computes a route for the requests revealed so far at the phase horizon, executes that route, and subsequently returns to the depot.
We further extend the phase-based capped-route framework of $\cite{BienkowskiKL21-ICALP}$). 


Related progress was made by Hwang and Jaillet \cite{Hwang}, who formulated online TRP within a broader framework of online scheduling with multi-state machines. Their framework yields a deterministic $5.14$-competitive deterministic algorithm.
This work also uses the phase-based framework. 

For the special case of line metrics, Bienkowski and Liu \cite{BienkowskiL19-MFCS} proposed a phase-based framework but modified computed routes to prioritize requests close to the depot. They obtained a deterministic $5.429$-competitive algorithm, an improvement   compared to $5.829$ for the arbitrary metric (which applies  to the lime metric). 
Their analysis combines a charging scheme with a factor-revealing linear program and exploits the geometry of the line. On a half-line (depot at the either end), Guragain and Sharma \cite{Guragain26} gave a deterministic $2+\sqrt{3}\approx 3.732$-competitive algorithm for 1-TRP. This bound, although seems promising and improved compared to 3.877, does not extend to the (full-)line.     


A significant improvement was obtained by Bienkowski, Kraska, and Liu \cite{BienkowskiKL21-ICALP} on which we build our framework. 
Compared to ours, their framework can be called {\em uninterruptible} since the entire route at each phase is executed without interruptions.
The competitive ratio of $4$ for $k$-TSP 
is established via an indirect method of formulating a complicated factor-revealing linear program and solving its dual. Our analysis avoids this via a direct cost accounting mechanism of bounding total virtual cost $B$ and physical cost $\mathrm{ALG}$,  and relating them to the optimal cost $\mathrm{OPT}$. 
With $k>1$, Bonifaci and Stougie \cite{bonifaci2009-TCS} gave the deterministic lower bound of $2$ for $k$-TRP. %





\vspace{2mm}
\noindent{\bf Organization of the Paper.} The remainder of the paper is organized as follows. We discuss details on the model and some preliminaries in Section \ref{sec:Model}. We then develop and analyze online $1$-TRP algorithm in Section \ref{sec:1-TRP}. We consider an extension to online $k$-TRP, $k\geq 2$, in Section \ref{sec:k-TRP}. The extensions to some variants of $k$-DARP, $k\geq 1$, are discussed in Section \ref{sec:k-DARP}. Finally, we conclude in Section \ref{sec:conclusion} with a short discussion on limitations of our work and future directions.

\begin{table}[ht!]
\footnotesize
\begin{center}
\resizebox{\textwidth}{!}{
\begin{tabular}{|p{2.2cm}|p{13.2cm}|}
\hline
\textbf{Variable} & \textbf{Description}  \\ \hline
$r=(a_r,x_r,w_r)$ & A TRP request with release time $a_r\geq 0$, location $x_r$, and weight $w_r>0$ \\ \hline
$o$ & Depot, located at $0$ \\ \hline
$d(x,y)$ & Distance between locations $x$ and $y$ \\ \hline 
$b_r$ & Earliest possible service time of request $r:\max\{a_r,d(0,x_r)\}$ \\ \hline
$m$ & Initial time scale: $m=\min_rb_r$ \\ \hline
$h_{j-1},h_{j},h_{j+1}$ & Phases $j-1$, $j$, and $j+1$ horizons 
\\ \hline 
$u$ & Elapsed time since the current phase horizon \\ \hline
$tc, q$ & Absolute scheduled cutoff time: $tc=(1+\alpha)h, \alpha\in (1/2,1), q=1+2\alpha$ \\ \hline
$P^*, P, P_j$ & An optimal route, a possible route, an optimal capped-route for phase $j$ 
\\
\hline
$C_r$ & Actual completion time of request $r$ by the online algorithm \\ \hline
$C^*_r$ & Optimal completion time of request $r$ by the optimal offline algorithm \\ \hline
$\mathrm{ALG}$ & Total weighted completion time of the online algorithm: $\sum_r w_rC_r$ \\ \hline
$\mathrm{OPT}$ & Minimum total weighted completion time of the optimal offline algorithm: $\sum_rw_rC^*_r$ \\ \hline

$c_{Pr}$ & Virtual service time of request $r$ under route $P$; $\infty$ if omitted \\ \hline
$c_{jr}$ & Virtual service time of request $r$ under route $P_j$ \\ \hline
$u_j$ & Actual duration of execution of $P_j$ 
\\ \hline
$\mathcal{L}(h_j)$ & 	Minimum capped-route cost in phase $j$ at cap $h_j: \mathcal{L}(h_j)=\min_P\sum_r w_r\min\{c_{jr},h_j\}$ \\ \hline
$\Delta_j$ & Increase in the optimal capped-route cost between caps $h_{j-1}$ and $h_j$: $\Delta_j=\mathcal{L}(h_j)-\mathcal{L}(h_{j-1})$ \\ \hline
$\mathcal{S}_j$ & Requests whose $c_{jr}$ are strictly before the current cap $h_j$ in phase $j$: $\mathcal{S}_j=\{r:c_{jr}<h_j\}$ \\ \hline
$\mathcal{D}$ & Requests registered in previous phases $1,2,\ldots, j-1$ \\ \hline
$\mathcal{A}_j$ & Requests eligible to be registered in phase $j: \mathcal{A}_j =\mathcal{S}_j\setminus D$  \\ \hline
$\mathcal{F}_j$ & Requests first-registered in phase $j: \{r\in \mathcal{A}_j:c_{jr}\le u_j\}$ \\ \hline
$w(\mathcal{X})$ & Total weight of a request set $\mathcal{X}: \sum_{r\in \mathcal{X}}w_r$ \\ \hline
$\mathcal{N}$ & New requests arrived after phase $j$ horizon: $\mathcal{N}=\{r:h_j<a_r\}$ \\ \hline
$\mathcal{V}_j$ & Requests in $\mathcal{A}_j$ not registered due to emergency cut in phase $j: \{r\in A_j: c_{jr}>t\}, h_j<t<tc$ \\ \hline
$\mathcal{T}_j$ & Requests in $\mathcal{A}_j$ not registered due to scheduled cut in phase $j$: $\{r\in A_j: c_{jr}>\alpha h_j\}$ \\ \hline
$\mathcal{Q}_j$ & Arrived requests not in $P_j$ at the scheduled checkpoint $tc: \{r:a_r\leq tc, r\notin \mathcal{D}\cup \mathcal{S}_j\}$ \\ \hline
$e_j$ & Virtual capped-route cost loss (residual cost) in phase $j$  caused by cutting $P_j$ \\ \hline
$\beta_{jr}$ & Virtual registration  cost of request $r$ (i.e., $r\in \mathcal{F}_j$) in phase $j: \beta_{jr}=\max\{h_{j-1},c_{jr}\}$ \\ \hline
$B$ & Total weighted virtual registration cost: $B=\sum_j\sum_{r\in \mathcal{F}_j}w_r\beta_{jr}$ \\ \hline
$K, M, \theta$ & Physical cost factor, emergency cutoff threshold, scheduled cutoff threshold \\ \hline 
$\mathcal{U}_j$ & Requests still not registered in phase $j$, including future arrivals \\ \hline
\end{tabular}
}
\end{center}
\vspace{-5mm}
\caption{Important variables used in Sections \ref{sec:Model} and \ref{sec:1-TRP} and their descriptions}
\label{table:variable}
\vspace{-3mm}
\end{table}

\section{Model and Preliminaries}
\label{sec:Model}
For simplicity in the exposition, we give model details and preliminaries we use in the 1-TRP algorithm (Section \ref{sec:1-TRP}).  Details related to $k$-TRP and $k$-DARP are given in Sections \ref{sec:k-TRP} and \ref{sec:k-DARP}. Table \ref{table:variable} lists the variables  we use to describe the model and preliminaries and the algorithm in Section \ref{sec:1-TRP}.  

\vspace{2mm}
\noindent{\bf Model.}
The server starts at the depot ($o=0$) at time $0$ and moves in (at most) unit speed. A finite set $\mathcal{R}$ of requests arrive over time. Each request $r\in \mathcal{R}$ is specified by a triple $r = (a_r, x_r,w_r)$, where $a_r \geq 0$ is its release time, $x_r\in \mathbb{R}^2$ is its location, and  $w_r > 0$ is its weight.
A request can be served only after its release. The completion
time $C_r$ of request $r$ is the first time at which the server visits $x_r$ at or after release time $a_r$, i.e., $C_r\geq a_r$. 
We assume that the server can move in any direction and can wait at any location. Service of a request is {\em instantaneous}.
We disregard requests released at the depot at time $0$, since they can be served immediately at zero cost. If no positive-cost requests remain, the problem is trivial. 
We assume that  $\mathcal{R}$ contains requests with arbitrary positive weights, arbitrary release times, and arbitrary locations in the metric space.

\vspace{2mm}
\noindent{\bf Online and Offline Algorithm Costs.}
The objective is to minimize the total weighted completion time
$\mathrm{ALG}=\sum_{r\in \mathcal{R}}
w_rC_r$ by an online algorithm. 
We denote the cost of an optimal offline algorithm by 
$\mathrm{OPT}=\sum_{r\in \mathcal{R}}w_rC^*_r$, where $C^*_r$ is the completion time of request $r$ by an optimal offline algorithm. 
The offline server has the same starting position and speed as the online server but knows the entire request sequence in $\mathcal{R}$ in advance. 

\vspace{2mm}
\noindent{\bf Metric Space.} Let $\mathcal{W}$ be the set of points and let $d:\mathcal{W}\times \mathcal{W}\rightarrow \mathbb{R}_{\geq 0}$ be a distance function. $\mathcal{W}$ admits an {\em arbitrary metric} if, for all distinct $x,y,z\in \mathcal{W}:$ (i) $d(x,y)\geq 0$, (ii) $d(x,y)=0$ if and only if $x=y$, (iii) $d(x,y)=d(y,x)$, and (iv) $d(x,z)\le d(x,y)+d(y,z)$. The word arbitrary means that we impose no additional restriction on the structure of the metric space. Therefore, the metric could be either a real line, Euclidean plane, or an undirected weighted graph (in which case the server moves along edges). The third property guarantees that the server traveling $u$ time units from $o$ can return to $o$ in $\leq u$ time units (the literature including \cite{BienkowskiKL21-ICALP} use this property). 

\vspace{2mm}
\noindent{\bf Route and Capped Route.} 
Let $\mathcal{R}({h_j})=\{r:a_r\leq h_j\}$ be the requests released until time $h_j$. 
Let $P$ be a route 
for the requests in $\mathcal{R}({h_j})$, which is a unit-speed walk starting at the depot $o$, possibly with waiting. 
We can represent $P$ as a trajectory $P:[0,l]\longrightarrow \mathcal{W}$ satisfying $P(0)=o$ and $d(P(s),P(t))\leq d(t,s),$ where $l$ is the length of $P$. Request $r$ is served when the server traversing $P$ visits $x_r\in P$ after its release time $a_r$.
Suppose the server has a cap of $h_j$ time units to traverse $P$. If $P$ contains $x_r$ at length $>h_j$, 
then $r$ is not served. 

\vspace{2mm}
\noindent{\bf Optimal Capped-Route Oracle.}
The route oracle computes a capped route $P$. 
Let one request-order be $r_1,r_2,\ldots r_n$. The oracle calculates the earliest feasible service times recursively: $c_1=\max\{a_1,d(o,x_1)\},$ $c_i=\max\{a_i,c_{i-1}+d(x_{i-1},x_i)\}.$ It then evaluates the following capped objective
$$\mathcal{L}(h_j)=\min_{P\in \mathcal{P}(\mathcal{R}(h_j))}\sum_r w_r\min\{c_{Pr},h_j\}$$
%
for each feasible request-order and chooses a request-order with the minimum cost (using deterministic tie-breaking if required). 
%
%
%
Here, $\mathcal{P}(\mathcal{R}(h_j))$ is the family of feasible  routes $P$ based on the requests revealed by time $h_j$ with cap $h_j$. 
This route oracle is {\em optimal} since
it finds a globally optimal route:
$$P_j\in \argmin_{P\in \mathcal{P}(\mathcal{R}(h_j))}\sum_rw_i\min\{c_{Pr},h_j\}.$$
$P_j$ minimizes the weighted completion times of requests in $\mathcal{R}(h_j)$, but capping the contribution of every request to  $h_j$. In any feasible route $P$, we have the following definition for $c_{Pr}$: 
\[
c_{P r} =
\begin{cases}
\text{service time of request }r \text{following}P, & \text{if served},\\
+\infty, & \text{if omitted}.
\end{cases}
\]

\vspace{2mm}
\noindent{\bf Initial Phase Horizon.}
Our framework works in phases of geometrically increasing lengths. We denote by {\em horizon} the start time of phase $j\geq 0$. Let $h_j$ denote the horizon associated with phase $j$. We also have, for each phase $j$, the cap of $h_j$ (equal to horizon $h_j$). 
These phases need to be initialized with initial reference horizon $h_0$. We initialize $h_0=m=\min_{r\in \mathcal{R}}b_r,$ with $b_r=\max\{a_r,d(o,x_r)\}.$   
The value $m$ can be learned by the online algorithm as follows: the server waits at the depot while maintaining the smallest value of $b_r$ among requests observed so far. 
By time $m$, every request $r$ with $b_r\leq m$ must have been released, since $a_r\leq b_r$. Thus the minimum value $m$ is known by the online algorithm by that time.  
For the finite set $\mathcal{R}$, 
every feasible route $P$ starting at the depot satisfies
$c_{Pr}\geq b_r\geq m,$ since  the server starts traversing $P$ at time $t>m$ from the depot; $c_{Pr}$ denotes the time at which the server serves $r$ following $P$.   
Therefore, it follows that for phase $0$,
$L(m)=m\sum_{r\in\mathcal{R}}w_r.$
Precisely, in our framework,
phase $1$ starts at horizon $h_1= (1+2\alpha) h_0$, $\alpha\in (1/2,1)$. The server starts to move from $o$ for the first time at $h_1$. Therefore, the server stays idle at $o$ during $[0,(1+2\alpha) h_0]$. 

\vspace{2mm}
\noindent{\bf Elapsed and Absolute Time.} In phase $j$, the {\em elapsed time} $u$ denotes the amount of time that has passed since $h_j$. If the current time is $t>h_j$, then  $u=t-h_j$. 
Notice that the server starts traversing $P_j$ at time $h_j$. 
Therefore, at the elapsed time $u$ in phase $j$, the server has traversed $u$ units of $P_j$. The {\em absolute time} (or simply time) denotes the amount of time that has passed since time $0$. In phase $j$, the absolute time related to the elapsed time $u$ is $h_j+u$. 

\vspace{2mm}
\noindent{\bf Virtual and Actual Service Time.}
In phase $j$, the {\em virtual service time} $c_{jr}$ denotes the elapsed time since $h_j$ at which request $r$ would be served by the server traversing  $P_j$ computed at the phase-$j$ horizon $h_j$. 
The {\em actual service time} corresponding to $c_{jr}$ is the absolute time $h_j+c_{jr}$.

\vspace{2mm}
\noindent{\bf Phase Specific Notation.}
For consecutive phases $j-1$ and $j$, let $\Delta_j$ denote the capped-cost increment:
$\Delta_j=\mathcal{L}(h_j)-\mathcal{L}(h_{j-1}).$
$\mathcal{L}$ is non-decreasing and hence $\Delta_j\geq 0$.
Let $P_j$ be the optimal capped route computed at horizon $h_j$ with cap $h_j$. 
Define
$\mathcal{S}_j=\{r:c_{jr}<h_j\}.$
Thus, $\mathcal{S}_j$ contains precisely the requests receiving a strictly positive benefit from being included in $P_j$.

Request $r$ is called {\em registered} in phase $j$ if its location $x_r$ is included in $P_j$ with cap $h_j$.
It is called {\em first-registered} in phase $j$ if the server visits $x_r$ while traversing $P_j$. 
Let $\mathcal{D}$ denote the set of requests first-registered in earlier phases (upto $j-1$), and define
$\mathcal{A}_j=\mathcal{S}_j\setminus \mathcal{D}.$
The set $\mathcal{A}_j$ consists of the candidates for first-registered in phase $j$.
If $P_j$ is executed for $u_j$ units of time, define its first-registered set by
$\mathcal{F}_j=\{r\in \mathcal{A}_j:c_{jr}\leq u_j\}.$
After phase $j$, the algorithm updates
$
\mathcal{D}\leftarrow \mathcal{D}\cup \mathcal{F}_j.$
$\mathcal{D}$ is kept fixed while the decisions within a phase are made.
At the elapsed time $u$, we denote by {\em tail} the trajectory of $P_j$ at the length interval $(u,h_j]$. 
If $P_j$ is cut at the elapsed time $u$, we denote by {\em unregistered tail} the requests in $\mathcal{A}_j$  that  could not be first-registered in phase $j$ since their locations belong to the tail.

For the analysis, we denote by  $\mathcal{U}_j$ the set of requests that remain {\em unregistered} in this bookkeeping sense after phase $j$, which are the requests that are not  first-registered until phase $j$ (i.e., not included in $\mathcal{D}$ yet). 
It includes requests not yet released. A request served incidentally during server traversing $P_j$  may already be completed, but it remains in the unregistered accounting until it becomes first-registered. 
We use $w(\mathcal{Z})=\sum_{r\in \mathcal{Z}}  w_r$ for the total weight of a request set $\mathcal{Z}$. 




\begin{algorithm}[bt!]
{
\footnotesize
{\bf Input:}

~~~~~Online requests $r = (a_r, x_r, w_r)$, one (at most) unit-speed server, depot $o = 0$






{\bf Initialization:}

~~~~~Handle requests released at $o$ at time 0; {\bf stop} if no positive-cost request remains
    
~~~~~{\bf for each} request $r$: $b_r = \max\{a_r, \mathrm{dist}(o,x_r)\}$, learn $m = \min_r b_r$ online by waiting at $o$

~~~~~$h_1 \leftarrow qm$, where $q=(1+2\alpha)$ for  $\alpha\in (1/2,1)$, $\mathcal{D} \leftarrow \emptyset$, parameters $M=\frac{\theta}{1-\alpha}$ and $\theta$

{\bf For phase $j = 1, 2, ...$:} 

~~~~~Set the absolute scheduled cutoff time for phase $j$: $tc = (1+\alpha)h_j$ 

~~~~~{\bf 1. CAPPED-ROUTE COMPUTATION}


~~~~~~~~~~~Compute an optimal route $P_j$  for all requests $\mathcal{R}(h_j)$ arrived until $h_j$ 
minimizing $\mathcal{L}(h)= \min_P \sum_r w_r \min\{c_{Pr}, h\},$ 

~~~~~~~~~~with $c_{Pr}$ be the virtual service time of request $r$ in route $P$

~~~~~~~~~~$c_{jr}\leftarrow$ the virtual service time of request $r$ in route $P_j$

~~~~~~~~~~$\mathcal{S}_j \leftarrow \{r : c_{jr} <  h_j\}, \mathcal{A}_j = \mathcal{S}_j \setminus \mathcal{D}$

~~~~~{\bf 2. CAPPED-ROUTE TRAVERSAL}

~~~~~~~~~~The server currently at $o$ traverses $P_j$ at time $h_j$ to serve requests 

~~~~~{\bf 3. EMERGENCY CHECKS: $h_j < t < tc$}

~~~~~~~~~~At every distinct time $t$, $h_j < t < tc$, at which new request(s) arrive:

~~~~~~~~~~~~~~~$u \leftarrow t - h_j, 
  \mathcal{N} = \{r : h_j < a_r \leq t\}, \mathcal{V} = \{r \in \mathcal{A}_j : c_{jr} > u\}$

~~~~~~~~~~{\bf if} $w(\mathcal{V}) > 0$ and $w(\mathcal{N}) \geq  M w(\mathcal{V})${\bf :} cut $P_j$ at time $t$, return directly to $o$, and set $h_{j+1}=qh_j=(1+2\alpha)h_j$ 



~~~~~{\bf 4. SCHEDULED CHECKS: $tc$}


~~~~~~~~~~$\mathcal{T} = \{r \in  A_j : c_{jr} > \alpha h_j\}$, $\mathcal{Q} = \{r : a_r \leq  tc \wedge r \notin \mathcal{D}\cup \mathcal{S}_j\}$

~~~~~~~~~~{\bf if}  $\mathcal{T} =\emptyset \vee w(\mathcal{Q}) \geq  \theta w(\mathcal{T})${\bf :} cut $P_j$ at time $tc$, return directly to $o$, and set $h_{j+1}=qh_j=(1+2\alpha)h_j$

~~~~~~~~~~{\bf else:} traverse $P_j$ fully until time $2h_j$, return directly to $o$, and set $h_{j+1} \leftarrow 3h_j$

~~~~~{\bf 5. REGISTRATION}

~~~~~~~~~~Let $h_j<u\leq 2h_j$ be the time at which server stops traversing $P_j$

~~~~~~~~~~$\mathcal{F}_j = \{r \in  A_j : c_{jr} \leq  u\}$, $D = D \cup  F_j$

~~~~~ {\bf if} requests remain to be served{\bf :}   start phase $j+1$ at time $h_{j+1}$ {\bf else:} stay at $o$


     \caption{Algorithm $Interruptible\_Capped\_Routes$}
\label{algorithm:capped}
}
\end{algorithm}

\begin{figure}[t] 
     \centering
         \includegraphics[width=0.9\textwidth]{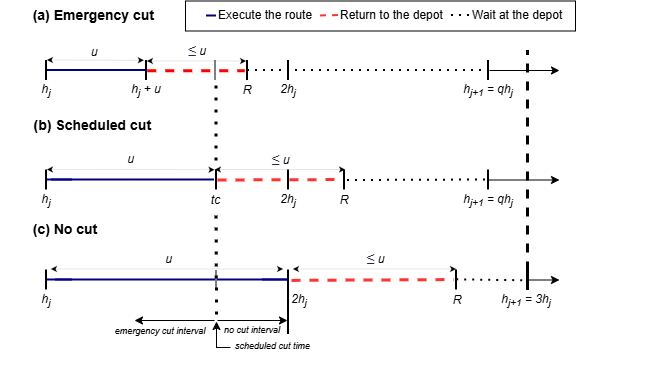}
   \vspace{-4mm}      
     \caption{An illustration of three possible outcomes of executing the capped route $P_j$ at phase-$j$ horizon $h_j$: ({\bf a}) An emergency cut stops executing $P_j$ at elapsed time $u<\alpha h_j, \alpha\in (1/2,1).$
({\bf b}) A scheduled cut stops executing $P$ at elapsed time $u=\alpha h_j$ (i.e.,  absolute time $tc=(1+\alpha)h_j$).
({\bf c}) No cut executes $P_j$ fully until elapsed time $u=h_j$.
Solid blue, dashed orange, and dotted black segments, respectively, denote the 
route execution, return to the depot, and waiting at the depot.
$R$ marks the arrival of the server at the depot.
Return durations are $\leq u$ for each of the three cases with their respective $u$ value.
Phase-$(j+1)$ horizon is the absolute time $h_{j+1}=qh_j=(1+2\alpha)h_j$ after either emergency or scheduled cut, and is $h_{j+1}=3h_j$
after no cut. An emergency cut can happen  only in the time interval $(h_j,(1+\alpha)h_j)$, the scheduled cut can only happen at time $(1+\alpha)h_j$. Additionally, in phase $j$, there will be at most a single cut (emergency or scheduled).
If no cut, then $P_j$ executes fully until absolute time $2h_j$ (the elapsed time $u=h_j$). 
}
     \label{fig:interruptible-routes}
     \vspace{-4mm}
\end{figure}

\section{$1$-TRP Algorithm}
\label{sec:1-TRP}
We now describe the online algorithm for $1$-TRP. 
The pseudocode is in Algorithm \ref{algorithm:capped}.  
Algorithm \ref{algorithm:capped} works in phases $j,j=1,2,\ldots,$ like  \cite{BienkowskiKL21-ICALP}. 
In the beginning of each phase $j$, the server is at depot $o$. During phase $j$, it traverses the trajectory $P_j$ calculated by the optimal capped-route oracle satisfying 
$P_j\in  \argmin_{P\in\mathcal{P}(\mathcal{R}_j)} \sum_r w_r \min\{c_{jr}, h_j\},$
with $h_j$ denoting the cap as well as horizon and $c_{jr}$ denoting the virtual service time $r$ following $P_j$. 

The main idea is that, our algorithm allows $P_j$ to be interruptible, i.e.,   in contrast to \cite{BienkowskiKL21-ICALP}, the server does not commit blindly to the entire $P_j$.
\begin{itemize}
\item If a large amount of new request weight arrives while it is traversing $P_j$, the algorithm cuts $P_j$ early and returns to the depot.
\item If $P_j$ has a large unregistered tail and many revealed requests lie outside that $P_j$, it cuts at the scheduled checkpoint and returns to the depot.
\item Otherwise, it finishes fully the phase and returns to the depot. The late visits in the tail provide enough accounting credit to pay for possible excess service time in the next phase.
\end{itemize}


Formally, let $\alpha\in (1/2,1)$. 
Let $tc$ denote the {\em scheduled check time} in phase $j$. 
In phase $j$, the server starts traversing $P_j$. 
During absolute time $(h,tc]$ decisions are made that will determine whether the server will cut $P_j$ and return directly to $o$. 
During $(h,tc)$, the decisions are made at each distinct time $t$ at which new request(s) arrive. These checks are called {\em emergency checks}. If no cut at time $t$,
the server simply follows $P_j$ from time $t$ until $t'>t$ at which new request(s) arrive.   
If all emergency checks during $(h,tc)$ do not cut $P_j$, then the last check in conducted exactly at time $tc$, which is the {\em scheduled check}. 
If the scheduled check at $tc$ also does not cut $P_j$, then no further checks are conducted during $(tc,2h_j]$ and hence the server traverses $P_j$ until absolute time $2h_j$ (or until the tail becomes empty). 
If $P_j$ was cut at the elapsed time $u$,  the server will return to the depot in next $\leq u$ time units. For an emergency cut $u<tc-h_j$, for a scheduled cut $u=tc-h_j$, and for no cut $u> tc$. Notice also that there is at most a single cut (emergency or scheduled) in a phase. 
In case of a cut, if server returns to $o$ before $qh_j=(1+2\alpha)h_j, q=(1+2\alpha)$, it waits at $o$ until  $qh_j$. 
In no cut,  if server returns to $o$ before $3h_j$, it waits at $o$ until $3h_j$. 
Fig.~\ref{fig:interruptible-routes} illustrates the above ideas. 

We now discuss how the emergency and scheduled cut decisions are made in phase $j$. These decisions play a crucial role in getting our claimed competitive ratio.  

\begin{itemize}
\item {\bf Emergency checks} are conducted at each distinct request arrival time $t$ with $h_j<t<tc$. Let $$\mathcal{N} = \{r: h_j < a_r \leq t\}, \mathcal{V} = \{r \in \mathcal{A}_j: c_{jr} > t\}.$$
The smallest $t$ at which   $w(\mathcal{V}) > 0$ and $w(\mathcal{N}) \geq  M w(\mathcal{V})$, the  server will cut $P_j$ and return directly to $o$. Here $M=\frac{\theta}{1-\alpha}$ with $\theta$ the scheduled cutoff threshold. 
\item {\bf Scheduled check} is conducted at absolute time $tc$.
Notice that the scheduled check happens in phase $j$ if and only if there is no emergency cut at phase $j$.   
Let $$\mathcal{T} = \{r \in  A_j : c_{jr} > tc\}, \mathcal{Q} = \{r : a_r \leq  tc \wedge r \notin \mathcal{D}\cup \mathcal{S}_j\}.$$
If  $\mathcal{T} =\emptyset$ or $w(\mathcal{Q}) \geq  \theta w(\mathcal{T})$, the server will cut $P_j$ and return to $o$.
Here $\theta$ is the scheduled cutoff threshold (will set value later). 
Since there is no check after $tc$ and if there is no scheduled cut at phase $j$, the server traverses $P_j$ until absolute time $2h_j$ and returns to $o$. %
\item {\bf Registration and return.}  
Let $u<h_j$ be the elapsed time at which server stops traversing $P_j$.
Let $\mathcal{F}_j = \{r \in  A_j: c_{jr} \leq  u\}$. We call the requests in $\mathcal{F}_j$ {\em first-registered} and add them to $\mathcal{D}$ (the requests first-registered during $[0,h_j+u]$.
The next phase starts at either $h_{j+1}=q h_j=(1+2\alpha)h_j$ (in case of emergency or scheduled cut) or $h_{j+1}=3h_j$ (no cut). 
The server stays at $o$ until $h_{j+1}$ and start the next phase $h_{j+1}$. 
Arrived requests during the server's return to $o$ are kept for phase $j+1$.
\end{itemize}
The server stays at $o$ after returning to $o$ when it finds that there is no request remains to be served. %

\subsection{Analysis of the Algorithm} 

We establish correctness of Algorithm \ref{algorithm:capped} as well as 3.877-competitive guarantee for any arbitrary instance of online 1-TRP. The analysis for the special class of zero-residual instances is done later. 

For any cap $h>0$ in $\mathcal{L}(h)$,
a request served after time $h$, or omitted from the route $P$, contributes $w_rh$. Consequently, increasing the cap cannot decrease the optimal capped objective:
$h'\geq h
\Longrightarrow
L(h')\geq L(h).
$
%
Indeed, evaluating the capped objective on an optimal offline route $P^*$ gives
$
L(h)
\leq
\sum_r w_r\min\{C_r^{*},h\}.$ 
%
Therefore, in the analysis, we may equivalently regard the oracle as optimizing over the full input, with unreleased requests assigned their constant capped contribution. We prove Lemma \ref{lemma:LhjOPT} which connects $\mathcal{L}(h_j)$ in each phase $j$ of Algorithm \ref{algorithm:capped} to the optimal offline cost $\mathrm{OPT}$. 

\begin{lemma}
\label{lemma:LhjOPT}
In each phase $j$ with horizon $h_j$ in Algorithm \ref{algorithm:capped},
    $\mathcal{L}(h_j) \leq \mathrm{OPT}.$
\end{lemma}
\begin{proof}
   Recall that $$\mathcal{L}(h_j)=\min_{P}\sum_rw_r\min\{c_{Pr},h_j\},$$ where $P$ is a feasible route and $c_{Pr}$ is the virtual service time of request $r$ under $P$. 
    Let $C^*_r$ be the completion time of request $r$ in the optimal offline route $P^*$. 
   Following $P^*$,
   by definition, $\mathrm{OPT}=\sum_rw_rC^*_r.$
   Now evaluate $P^*$ using the capped objective: $\sum_rw_r\min\{C^*_r,h_j\}.$ Since $\min\{C^*_r,h_j\}\leq C^*_r$, we have
$$\sum_rw_r\min\{C^*_r,h_j\}\leq \sum_rw_r C^*_r=\mathrm{OPT}.$$

 Considering $\mathcal{L}(h_j)$, it is the minimum capped-route cost over all feasible routes $P\in \mathcal{P}(\mathcal{R}(h_j))$. Therefore,
 $\mathcal{L}(h_j)\leq \sum_r \min\{C^*_r,h_j\}.$
 Combining the above two inequalities gives $\mathcal{L}(h_j)\leq \mathrm{OPT}$.
\end{proof}

We now establish the correctness of Algorithm \ref{algorithm:capped} -- each request in the finite $\mathcal{R}$ of requests are eventually registered by the server executing Algorithm \ref{algorithm:capped}.

\begin{lemma}
\label{lemma:serve}
    For a finite set $\mathcal{R}$ of requests, every request $r\in \mathcal{R}$ is eventually registered by Algorithm \ref{algorithm:capped} (i.e., included in set $\mathcal{F}_j$ of some phase $j$).
\end{lemma}
\begin{proof}
With $\alpha\in (1/2,1)$ and $q=1+2\alpha$,
the horizon $h_{j+1}$ for phase $j+1$ is chosen as follows:
\[
h_{j+1} = 
\begin{cases} 
      qh_j, & \text{after an emergency or scheduled cut} \\
      3h_j, & \text{after finishing the phase (i.e., no cut)} 
   \end{cases}
\]
Therefore, $h_{j+1}\geq q h_j.$ Since $q>2$ for $\alpha\in (1/2,1)$, the phase horizons grows geometrically. 

For the finite set $\mathcal{R}$ of requests, define $w_{\min}=\min_{r\in\mathcal{R}}w_r>0.$
We assume that requests at the depot $o$ released at time 0 have already been registered. Thus, if any requests remain, $\mathrm{OPT}>0.$ Because $\mathcal{R}$ is finite, there is the maximum  release time
$a_{\max}=\max_{r\in \mathcal{R}}a_r.$
We will show that a sufficiently late phase registers every remaining request. Since $q>2$, $\lim_{j\rightarrow\infty}{h_j}=\infty.$ Consequently, we can choose a phase $j$ sufficiently large such that both $h_j\geq a_{\max}$ and $\alpha h_j>\frac{\mathrm{OPT}}{w_{\min}}.$

Let us examine phase $j$. Since $h_j\geq a_{\max}$, all requests in $\mathcal{R}$ must have been released by the phase-$j$ horizon $h_j$. The oracle computes an optimal capped route with cap $h_j$: 
$$\mathcal{L}(h_j)=
\min_P\sum_{r\in \mathcal{R}}
w_r\min\{c_{Pr},h_j\}.$$
From Lemma \ref{lemma:LhjOPT}, we know that $\mathcal{L}(h_j)\leq \mathrm{OPT}.$
Let $P_j\in \mathcal{P}(\mathcal{R}(h_j))$ be the selected optimal route and let $c_{jr}$ be the virtual service time of request $r$ in $P_j$. 
We have that $w_r\min\{c_{jr},h_j\}\leq \mathrm{OPT}.$
Dividing by $w_r$, $\min\{c_{jr},h_j\}\leq \mathrm{OPT}/w_r$.
Since $w_r\geq w_{\min}$, $\min\{c_{jr},h_j\}\leq \mathrm{OPT}/w_{\min}$.
Now recall that the elapsed time for the scheduled cutoff at phase $j$ is  $\alpha h_j.$ Since $\alpha h_j>\frac{\mathrm{OPT}}{w_{\min}}$, $\frac{\mathrm{OPT}}{w_{\min}}<\alpha h_j<h_j.$
Combining this with $\min\{c_{jr},h_j\}\leq \mathrm{OPT}/w_{\min}$, we get $c_{jr}<\alpha h_j$ for every request $r\in \mathcal{R}$. In particular, $\mathcal{S}_j=\mathcal{R}.$

We now show that an emergency cut cannot occur during phase $j$.  The emergency rule checks new requests arriving at time $t>h_j$: $\mathcal{N}(t)=\{r:h_j<a_r\le t\}.$
But all requests have already been released by time $h_j$.
Therefore, $\mathcal{N}(t)=\emptyset$
throughout phase $j$. Hence the emergency condition
$w(\mathcal{N}(t))\ge M w(\mathcal{V}_j(u))$
cannot trigger. 
So the algorithm reaches the scheduled checkpoint $tc=(1+\alpha)h_j$ in phase $j$. 
At the schedule checkpoint, the algorithm computes $\mathcal{T}_j=\{r\in \mathcal{A}_j: c_{jr}>\alpha h_j\}.$ But we have shown that every request satisfies $c_{jr}<\alpha h_j$. Therefore, $\mathcal{T}_j=\emptyset$. According to the algorithm, $\mathcal{T}_j=\emptyset$ always triggers a scheduled cut at the elapsed time $\alpha h_j$. The served requests in phase $j$ are $\mathcal{F}_j=\{r\in \mathcal{A}_j:c_{jr}\leq \alpha h_j\}.$ Since every request in $\mathcal{A}_j$ has $c_{jr}<\alpha h_j, \mathcal{F}_j =\mathcal{A}_j$. Recall that $\mathcal{A}_j=\mathcal{S}_j\setminus \mathcal{D}$. Since $\mathcal{S}_j=\mathcal{R}, \mathcal{F}_j=\mathcal{R}\setminus \mathcal{D}$. Thus, every request not registered in an earlier phase is registered in phase $j$. The updated registered set becomes $\mathcal{D}\leftarrow \mathcal{D}\cup \mathcal{F}_j=\mathcal{D}\cup (\mathcal{R}\setminus \mathcal{D})=\mathcal{R}.$
Therefore, every request $r\in \mathcal{R}$ has been registered in phase $j$. The lemma follows.
\end{proof}

We now quantify the capped-route cost  loss (residual cost), denoted by $e_j$, caused by cutting $P_j$  in phase $j$ either through the emergency cut or the scheduled cut. 
Whatever the value of $e_j$ is, it needs to be added to the increments in the capped offline optimal costs $\Delta_j=\mathcal{L}(h_j)-\mathcal{L}(h_{j-1})$. This addition increases the cost online algorithm pays w.r.t. $\mathrm{OPT}$ as  $OPT\geq \mathcal{L}(h_j)$ (Lemma \ref{lemma:LhjOPT}) since $\Delta_j$ is related to $\mathcal{L}(h_j)$. Therefore, we need to guarantee that $e_j$ is small compared to $\Delta_j$ so that $\Delta_j+e_j$ remains closer to $\Delta_j$ and hence closer to $\mathrm{OPT}$.


Notice that if $P_j$ is run until cap $h_j$ (i.e., no cut), then $e_j=0$ as no benefit is lost -- all requests that $P_j$ was planning to register in phase $j$ are registered.    
However, if $P_j$ is cut before cap $h_j$, then some benefit is lost which $e_j$ quantifies. 
To put it another way, we can think of $e_j$ as the price of abandoning part of $P_j$.  If $P_j$ is abandoned at time $t>h_j$ closer to horizon $h_j$, the loss is large giving large $e_j$. However, if $P_j$ is abandoned at time $t>h_j$ closer to the cap $h_j$, the loss is small giving small $e_j$. This is the reason our algorithm cuts $P_j$ either sufficiently late or there is enough weight outside $P_j$ to justify the early interruption (decided through parameters $M$ and $\theta$). 
We have Lemma \ref{lemma:ej} for $e_j$ at phase $j$.




\begin{lemma}
\label{lemma:ej}
In any phase $j$ in Algorithm \ref{algorithm:capped},
$$e_j=\sum_{r\in \mathcal{A}_j\setminus \mathcal{F}_j} w_r(h_j-c_{jr}).$$
\end{lemma}
\begin{proof}
Recall $\mathcal{A}_j=\mathcal{S}_j\setminus \mathcal{D}$, where $\mathcal{S}_j=\{r:c_{jr}<h_j\}$ consists of requests that the $P_j$ plans to register before the cap $h_j$. Let $\mathcal{F}_j=\{r\in A_j: c_{jr}\leq u\}$ be the requests that are registered during phase $j$ before $P_j$ is cut at the elapsed time $u$. Therefore, $\mathcal{A}_j\setminus \mathcal{F}_j$ contains requests that $P_j$ planned to register before the cap $h_j$, but they were not registered due to the cut. Precisely the requests in the set $\mathcal{A}_j\setminus \mathcal{F}_j$ contribute to $e_j$. 

Consider one such request $r$ such that $x_r$ is in  $P_j$. Its capped contribution is $w_rc_{jr}$ because $c_{jr}<h_j$. 
If $P_j$ is cut before $r$ is registered, its capped contribution becomes $w_rh_j$. 
Thus $r$ contributes $w_r(h_j-c_{jr})$ to the capped-route cost loss. 
Therefore, the total $e_j$ at phase $j$ is the sum of the contributions due to all the requests in $\mathcal{A}_j\setminus \mathcal{F}_j$, which is $\sum_{r\in \mathcal{A}_j\setminus \mathcal{F}_j} w_r(h_j-c_{jr}).$
\end{proof}

We now relate $e_j$ to $\Delta_j=\mathcal{L}(h_j)-\mathcal{L}(h_{j-1})$, 
the increase in the optimal capped-route cost between subsequent caps $h_{j-1}$ and $h_j$. 
We prove the following lemma.
\begin{lemma}
\label{lemma:ejdeltaj}
Let $\theta=\frac{(2\alpha-1)(1+\alpha)}{2(1-\alpha)}$ and $M=\frac{\theta}{1-\alpha}$, then in any phase $j$ of Algorithm \ref{algorithm:capped},
$e_j\leq \frac{q}{M(q-1)}\Delta_j.$
\end{lemma}
\begin{proof}
    At phase $j$, consider the set of requests in $\mathcal{Q}_j$ that are not in the route $P_j$ with cap $h_j$. At phase $j$, each $r\in \mathcal{Q}_j$ contributes $w_rh_j$. If we evaluate $P_j$ with a smaller cap $h_{j-1}$, $r$ contributes only $w_rh_{j-1}.$ This can be done in Lemma \ref{lemma:ej} replacing $c_{jr}$ by $h_{j-1}$ since $c_{jr}>h_{j-1}$. This replacement facilitates us to compare $e_j$ with $\Delta_j$ since $\Delta_j$ depends on $h_{j-1}$ not $c_{jr}$. Therefore, the total reduction is at most $\sum_{r\in\mathcal{Q}_j}w_r(h_j-h_{j-1})=(h_j-h_{j-1})w(\mathcal{Q}_j).$ Since $P_j$ is the optimal capped route for phase $j$, even setting $h_{j-1}$ instead of $c_{jr}$ in Lemma \ref{lemma:ej}, the total reduction cannot exceed the corresponding increase in the optimal capped-route cost $\Delta_j$  between two caps $h_{j-1}$ and $h_{j}$. Therefore,
    $$(h_j-h_{j-1})w(\mathcal{Q}_j)\leq \Delta_j.$$
    Using the minimum growth factor $q$, $h_j\geq qh_{j-1}$ for $q>2$. Rearranging, $h_j-h_{j-1}\geq h_j-\frac{h_j}{h_{j-1}}=\frac{q-1}{q}h_j.$ Hence,
    \begin{align}
        \frac{h_j}{h_j-h_{j-1}}\leq \frac{q}{q-1}.
    \end{align}
    We now compare the capped-route cost loss  $e_j$ at phase $j$ and the optimal capped cost increase $\Delta_j$ between phases $j$ and $j-1$.
    We consider two cases:
    \begin{itemize}
        \item {\bf Case 1: Emergency cut of $P_j$ at absolute time $h_j<t<tc=(1+\alpha)h_j$.} If Algorithm \ref{algorithm:capped} cuts $P_j$ then new requests $\mathcal{N}$ have arrived in phase $j$, with $w(\mathcal{N})\geq M w(\mathcal{V}_j)$, where $\mathcal{V}_j$ is requests in $\mathcal{A}_j$ not registered due to emergency cut in phase $j$ at time $h_j<t<tc=(1+\alpha)h_j$. Due to this emergency cut, $e_j\leq h_j w(\mathcal{V}_j)$. Using the emergency cut condition, $e_j\leq \frac{h_j}{M} w(\mathcal{N})$.
        Set $\mathcal{Q}_j=\mathcal{N}$ in $(h_j-h_{j-1})w(\mathcal{Q}_j)\leq \Delta_j$. 
        Which gives $w(\mathcal{N})\leq \frac{\Delta_j}{h_j-h_{j-1}}.$
        Therefore, $$e_j\leq \frac{h_j}{M} w(\mathcal{N})\leq \frac{h_j}{M(h_j-h_{j-1})} \Delta_j.$$
        Finally, applying Equation 1, 
        $e_j\leq \frac{q}{M(q-1)} \Delta_j.$
        \item {\bf Case 2: Scheduled cut of $P_j$ at absolute time $tc=(1+\alpha)h_j$.} Algorithm \ref{algorithm:capped} cuts $P_j$ at time $tc=(1+\alpha) h_j$ only when $w(\mathcal{Q}_j)\geq \theta w(\mathcal{T}_j)$, where $\mathcal{T}_j=\{r\in A_j: c_{jr}>\alpha h_j\}$. Each $r\in \mathcal{T}_j$ has service time greater than $c_{jr}>\alpha h_j$. Therefore, $h_j-c_{jr}<(1-\alpha)h_j$. The total capped-route cost loss satisfies: $e_j\leq (1-\alpha)h_jw(\mathcal{T}_j).$
        The scheduled cut condition gives $w(\mathcal{T}_j)\leq \frac{1}{\theta} w(\mathcal{Q}_j).$
        Thus, $e_j\leq \frac{(1-\alpha)h_j}{\theta}w(\mathcal{Q}_j).$
        Choosing $M=\frac{\theta}{1-\alpha}$ so that the emergency and scheduled interruption cost bounds become equal, we obtain
        $e_j\leq \frac{h_j}{M}w(\mathcal{Q}_j).$
        Since $\mathcal{Q}_j$ is outside $\mathcal{S}_j$, 
        applying  $(h_j-h_{j-1})w(\mathcal{Q}_j)\leq \Delta_j$,
        $$e_j\leq \frac{h_j}{M} w(\mathcal{N})\leq \frac{h_j}{M(h_j-h_{j-1})} \Delta_j \leq \frac{q}{M(q-1)} \Delta_j.$$
    \end{itemize}
     If phase $j$ has no interruption or an empty tail, 
     $e_j=0$, so the inequality holds trivially.
%
\end{proof}

We now consider $e_j$ for all the phases 
and connect with the optimal offline cost $\mathrm{OPT}$. We first need to connect $\sum_j\Delta_j$ with $\mathrm{OPT}.$

\begin{lemma}
\label{lemma:DeltaOPT}
Summing over phases until  phase $j$ in Algorithm \ref{algorithm:capped},
    $\sum_j\Delta_j\leq   \mathrm{OPT}.$
\end{lemma}
\begin{proof}
 We have that 
    $$\sum_j\Delta_j=\sum_j \left(\mathrm{L}(h_j)-\mathrm{L}(h_{j-1})\right)=\mathcal{L}(h_j)-\mathcal{L}(h_0)=\mathcal{L}(h_j)-\mathcal{L}(m).$$
From  Lemma \ref{lemma:LhjOPT}, $\mathcal{L}(h_j)\leq \mathrm{OPT}$. 
Since $\mathcal{L}(m)>0, \sum_j\Delta_j\leq \mathcal{L}(h_j).$
Therefore, $\sum_j\Delta_j\leq \mathrm{OPT}$. 
\end{proof}
\begin{lemma}
\label{lemma:ejOPT}
Summing over phases until phase $j$ in Algorithm \ref{algorithm:capped},
    $\sum_je_j\leq  \frac{q}{M(q-1)} \mathrm{OPT}.$
\end{lemma}
\begin{proof}
    From Lemma \ref{lemma:ejdeltaj}, in phase $j$, $e_j\leq \frac{q}{M(q-1)}\Delta_j.$
    Summing over all phases and using Lemma \ref{lemma:DeltaOPT},  
    $$\sum_je_j\leq \frac{q}{M(q-1)} \sum_j\Delta_j\leq \frac{q}{M(q-1)} \mathrm{OPT}.$$ 
\end{proof}

We now bound $B$: total weighted virtual registration cost of all requests, counted exactly once when they become first-registered.  
Note that a request $r$ accumulates credit while it remains unregistered. Once it is registered, its accumulated credit becomes exactly $w_r\max\{h_{j-1},c_{jr}\}$, which is $r$'s contribution to $B$. The capped-route cost loss $e_j$ due to interruptions  compensate for the accumulated credit lost when $P_j$ is  cut. Thus, summing over phases accounts for every request exactly once. 
To make analysis simpler, we charge $e_j$ to $\Delta_j$ to see how much $\mathrm{OPT}$ increases due to interruptions. Regarding $\mathrm{ALG}$, if  phase $j-1$ was interrupted, phase $j$ starts early and that gives excess in $\mathrm{ALG}$ which we charge to unused allowance from the unregistered tail of $P_{j-1}$. 
Therefore, the bound $B$ is important and heart of our analysis since it serves as the connection between $\mathrm{ALG}$  and $\mathrm{OPT}$ as follows. We establish the following connections in Lemmas \ref{lemma:BOPT} and \ref{lemma:KB}:

\[
\underbrace{\text{optimal offline cost}}_{OPT}
\quad\overset{\text{Lemma \ref{lemma:BOPT}}}{\longrightarrow}\quad
\underbrace{\text{total virtual registration cost}}_{B\leq \left(1+\frac{q}{M(q-1)}\right)OPT} \quad\overset{\text{Lemma \ref{lemma:KB}}}{\longrightarrow}\quad
\underbrace{\text{online algorithm cost}}_{ALG \leq KB}.
\]

\begin{lemma}
\label{lemma:B}
Summing over phases until phase $j$ in Algorithm \ref{algorithm:capped},
    $$B=\sum_j\sum_{r\in \mathcal{F}_j}w_r\beta_{jr},$$ where $\beta_{jr}=\max\{h_{j-1},c_{jr}\}.$
\end{lemma}
\begin{proof}
    Recall that $\mathcal{F}_j=\{r\in A_j: c_{jr}\leq u\}$ are the requests first-registered during phase $j$ before $P_j$ is cut at the elapsed time $u$. Note that every request belongs to exactly one $\mathcal{F}_j$, namely the phase in which it is first-registered.
    We first prove $\beta_{jr}=\max\{h_{j-1},c_{jr}\}.$
    We consider two cases.

    \begin{itemize}
    \item {\bf Case 1: $c_{jr}<h_{j-1}$.} In this case $\beta_{jr}=h_{j-1}$. In other words, even though $P_j$ registers the request early, the algorithm does not begin executing $P_j$ until time $h_j$. Therefore, the previous phase horizon is the relevant baseline for its first-registered cost.
    \item {\bf Case 2: $c_{jr}\geq h_{j-1}$.} In this case, $\beta_{jr}=c_{jr}$. In other words, the virtual service time itself determines the first-registered cost.
    \end{itemize}

    We now prove $B=\sum_j\sum_{r\in \mathcal{F}_j}w_r\beta_{jr}$. Consider a request $r$ that is first-registered in phase $j$. Before phase $1$, its credit is $m w_r$. In the subsequent phases until phase $j-1$, the total is 
    $w_r(h_1-h_0)+\ldots+w_r(h_{j-1}-h_{j-2}).$
    Finally, in phase $j$, the increment is $\max\{w_r(c_{jr}-h_{j-1}),0\}.$ 
    Adding everything, 
    $$mw_r+w_r(h_{j-1}-m)+\max\{w_r(c_{jr}-h_{j-1}),0\}=w_r\beta_{jr}.$$
    Therefore, considering all the requests first-registered until phase $j$, $B=\sum_j\sum_{r\in \mathcal{F}_j}w_r\beta_{jr}.$
\end{proof}

We now relate $B$ to $\mathrm{OPT}$. 

\begin{lemma}
\label{lemma:BOPT}
Given the optimal route $P_j$, until phase $j$,
$$B\leq \mathrm{OPT}+\sum_j e_j\leq \left(1+\frac{q}{M(q-1)}\right)OPT.$$
\end{lemma}
\begin{proof}
Since $P_j$ is optimal at cap $h_j$,
$\mathcal{L}(h_j)= \sum_r w_r \min\{c_{jr}, h_j\}.$
Evaluate $P_j$ in the previous cap $h_{j-1}:  \sum_r w_r \min\{c_{jr}, h_{j-1}\}.$
But $P_j$ might not be optimal at $h_{j-1}$. Therefore,
$\mathcal{L}(h_{j-1})\leq \sum_r w_r \min\{c_{jr}, h_{j-1}\}.$
Subtracting gives $$\Delta_j:=\mathcal{L}(h_j)-\mathcal{L}(h_{j-1})\geq \sum_r w_r \min\{c_{jr}, h_j\}-\sum_r w_r \min\{c_{jr}, h_{j-1}\}.$$
Let $\delta_{jr}=w_r \min\{c_{jr}, h_j\}-w_r \min\{c_{jr}, h_{j-1}\}$ denote the increment for individual request $r$. 
This gives $\Delta_j\geq \sum_r \delta_{jr}.$
\begin{itemize}
\item {\bf Case A: Request $r$ is registered in phase $j$.} That means $r\in \mathcal{F}_j$ and hence $c_{jr}<h_J$. Therefore, 
$$\delta_{jr}=w_r c_{jr}-\min\{c_{jr}, h_{j-1}\}=w_r\max\{c_{jr}-h_{j-1},0\}.$$
The is the credit earned by $r$ that is registered in the current phase $j$.
\item {\bf Case B: Request $r$ is not registered in phase $j$ and not in $P_j$.} That means $r\in \mathcal{U}_j\setminus \mathcal{S}_j.$ Since $r\notin \mathcal{S}_j$, its virtual service time satisfies $c_{jr}\geq h_j$.
Consequently, $$\delta_{jr}=w_r(h_j-h_{j-1}).$$ This means $r$ contributes the full cap increment. 
\item {\bf Case C: Request $r$ is in $P_j$ but $P_j$ was cut before $r$ is registered.} That means $r\in \mathcal{U}_j \cap \mathcal{S}_j=\mathcal{A}_j\setminus \mathcal{F}_j.$ Since $c_{jr}<h_j$ for $r$ but not registered in phase $j$, its increment $\delta_{jr}=w_r \max\{c_{jr}-h_{j-1},0\}$. Additionally, the capped-route cost loss $e_j$ due to interruption is $w_r(h_j-c_{jr})$. 
Adding these two quantities
$$\delta_{jr}+w_r(h_j-c_{jr})=w_r \max\{c_{jr}-h_{j-1},0\}+w_r(h_j-c_{jr})\geq w_r(h_j-h_{j-1}),$$ because $\max\{c_{jr}-h_{j-1},0\}\geq c_{jr}-h_{j-1}.$
\end{itemize}
Combining the costs in Cases A, B, and C, we obtain for phase $j$,
$$\Delta_j+e_j\geq (h_j-h_{j-1})w(\mathcal{U}_j)+\sum_{r\in \mathcal{F}_j} w_r\max\{c_{jr}-h_{j-1},0\}.$$

We now sum over all the phases $1,2,\ldots,j$, which gives
\begin{equation}
\sum_{j} (\Delta_j+e_j) \geq \sum_j (h_j-h_{j-1})w(\mathcal{U}_j)+\sum_j \sum_{r\in \mathcal{F}_j} w_r\max\{c_{jr}-h_{j-1},0\}.    
\end{equation}

We now consider an arbitrary request $r$ which is first-registered in phase $j$ (meaning $r\in \mathcal{F}_j$) and determine its contribution to $\sum_{j} (\Delta_j+e_j)$.
Observe that at $h_0=m$, request $r$ receives credit $\mathcal{L}(m)=m\sum_rw_r$. This is because, every request has earliest possible service time at least $m$ and hence, at cap $m$, every request contributes  exactly $w_rm$, regardless of the route. 

Before phase $j$, $r$ was not registered. Thus, $r\in U_i,i<j.$ 
Therefore, in each phase $k<j$, $r$ contributes $(h_k-h_{k-1})w_r.$ 
The accumulated credit for $r$ until phase $j-1$  is 
$$\geq w_r\left(m+\sum_{i=1}^{j-1}(h_i-h_{i-1})\right)=mw_r+w_r(h_{j-1}-m).$$ 
In phase $j$, since $r\in \mathcal{F}_j$ (and hence first-registered), it receives an additional credit $w_r\max\{c_{jr}-h_{j-1},0\}.$
Adding all the credits received by request $r$, we obtain
$$mw_r+w_r(h_{j-1}-m)+w_r\max\{c_{jr}-h_{j-1},0\}=w_r\max\{h_{j-1},c_{jr}\}=w_r\beta_{jr}.$$
This is exactly the contribution of request $r$ to $B$.

We now repeat this for each request. Notice that each request is eventually registered (Lemma \ref{lemma:serve}). Moreover, each request is first-registered in exactly one phase.  
Consequently, when we sum the credit contributions over all requests (once all requests have been registered), we obtain
\begin{align}
    \mathcal{L}(m)+\sum_j (h_j-h_{j-1})w(\mathcal{U}_j)+\sum_j \sum_{r\in \mathcal{F}_j} w_r\max\{c_{jr}-h_{j-1},0\} &= \sum_j \sum_{r\in \mathcal{F}_j} w_r\max\{c_{jr},h_{j-1}\} \\
    &=\sum_j\sum_{r\in \mathcal{F}_j} w_r\beta_{jr} = B.
\end{align}

Substituting $\sum_j (h_j-h_{j-1})w(\mathcal{U}_j)+\sum_j \sum_{r\in \mathcal{F}_j} w_r\max\{c_{jr}-h_{j-1},0\}$ by $\sum_{j} (\Delta_j+e_j)$ (Equation 2), we obtain
$$\mathcal{L}(m)+\sum_{j} (\Delta_j+e_j)\geq B.$$
This can be written as 
$$B\leq \mathcal{L}(m)+\sum_{j} \Delta_j+\sum_j e_j.$$
Using $\sum_j\Delta_j=\mathcal{L}(h_j)-\mathcal{L}(m)$ (Lemma \ref{lemma:DeltaOPT}) and $\mathcal{L}(h_j)\leq \mathrm{OPT}$ (Lemma \ref{lemma:LhjOPT}), we obtain 
$$B\leq \mathrm{OPT}+\sum_je_j.$$

Now plugging the value of $\sum_je_j$ from Lemma \ref{lemma:ejOPT}, we obtain
$$B\leq \mathrm{OPT}+\sum_je_j\leq \left(1+\frac{q}{M(q-1)}\right)OPT.$$
\end{proof}

Finally, we now relate $\mathrm{ALG}$ to $B$ (which in turns relates $\mathrm{ALG}$ to  $\mathrm{OPT}$).

\begin{lemma}
\label{lemma:KB}
    With $K=1+q,q=1+2\alpha, \alpha\in (1/2,1)$,
    $$\mathrm{ALG}\leq \sum_j\sum_{r\in \mathcal{F}_j}w_r(h_j+c_{jr})\leq KB.$$
\end{lemma}
\begin{proof}
    Consider phase $j$ and a request $r\in\mathcal{F}_j$. The server starts phase $j$ from the depot at absolute time $h_j$. It then traverses the route $P_j$. The virtual service time for $r$ in $P_j$ is $c_{jr}$. Therefore, the server visits $r$ at absolute time $H_{jr}=h_j+c_{jr}.$
    The request $r$ may even be registered earlier, for example, if the server encounters its location $x_r$ incidentally. Thus,
    $$C_r\leq h_j+c_{jr}=H_{jr}.$$
    Each request is first-registered exactly once, so
    $\sum_j\sum_{r\in \mathcal{F}_j}w_rH_{jr}.$
    It remains to prove $$\sum_j\sum_{r\in \mathcal{F}_j}w_rH_{jr}\leq KB.$$
    Equivalently, we want to establish 
    $$\sum_j\sum_{r\in \mathcal{F}_j}w_r(H_{jr}-K\beta_{jr})\leq 0,$$
    since from Lemma \ref{lemma:B}, $B=\sum_j\sum_{r\in \mathcal{F}_j}w_r\beta_{jr}.$ 
    
    We consider two cases:
    \begin{itemize}
        \item {\bf Case 1: The preceding phase $j-1$ was cut.} Suppose phase $j-1$ was cut, either by the emergency rule or the scheduled rule. 
        Phase $j$ starts at $h_j=qh_{j-1}.$ 
        For any request $r\in \mathcal{F}_j$, $H_{jr}=qh_{j-1}+c_{jr}$ and $\beta_{jr}=\max\{h_{j-1},c_{jr}\}.$
        We want to show 
        $$qh_{j-1}+c_{jr}\leq (q+1)\max\{h_{j-1},c_{jr}\}.$$
        These are two sub-cases.
        \begin{itemize}
            \item {\bf Case 1a: $c_{jr}\leq h_{j-1}$.} 
            In this case, $$H_{jr}=qh_{j-1}+c_{jr}\leq qh_{j-1}+h_{j-1}=(q+1)h_{j-1}=K\beta_{jr}.$$
            \item {\bf Case 1b: $c_{jr}>h_{j-1}$.} In this case, $$H_{jr}=qh_{j-1}+c_{jr}\leq qc_{jr}+c_{jr}=(q+1)c_{jr}=K\beta_{jr}.$$
        \end{itemize}
        Therefore, whenever the previous phase $j-1$ was interrupted, $H_{jr}\leq K\beta_{jr}.$  The same argument applies to phase $1$ because $h_1=qh_0, h_0=m$.
        Therefore, the requests first-registered in phase $1$ or in a phase following an interruption guarantee that $\sum_j\sum_{r\in \mathcal{F}_j}w_r(H_{jr}-K\beta_{jr})\leq 0.$
        
        Since the preceeding phase $j-1$ was cut, $h_j=qh_{j-1}$. Therefore, $qh_{j-1}+c\leq (q+1)\max\{h_{j-1},c\}=K\beta,$ since $K=1+q.$
        Phase $1$ also satisfies this inequality because $h_1=qm.$ Thus, every request first-registered following a cut is individually covered  in the calculation.
        \item {\bf Case 2: The preceding phase $j-1$ was not cut.} 
        Suppose phase $j-1$ had horizon $h_{j-1}$ and was not cut. The next phase $j$ starts at  $h_j=3h_{j-1}$.
        For a request $r\in \mathcal{F}_j$, denote its virtual service time by $c'$. Then $H_{jr}=3h_{j-1}+c'$ and $\beta_{jr}=\max\{h_{j-1},c'\}$. 
        Define its excess over the normal allowance: $\mathcal{E}(c')=3h_{j-1}+c'-K\max\{h_{j-1},c'\}.$
        We will show that $\mathcal{E}(c')\leq (3-q)h_{j-1}.$
        We again consider two sub-cases.
        \begin{itemize}
            \item {\bf Case 2a: $c'\leq h_{j-1}$.} Then 
            $$\mathcal{E}(c')=3h_{j-1}+c'-Kh_{j-1}=(3-K)h_{j-1}+c'.$$
            Since $c'\leq h_{j-1}$ and $K=1+q$,
            $$\mathcal{E}(c')=(3-K)h_{j-1}+h_{j-1}=(4-K)h_{j-1}=(3-q)h_{j-1}.$$
            \item {\bf Case 2b: $c'> h_{j-1}$.} Then 
            $$\mathcal{E}(c')=3h_{j-1}+c'-Kc'=3h_{j-1}-(K-1)c'=3h_{j-1}-qc'.$$
            Since $c'>  h_{j-1}$,
            $$\mathcal{E}(c')\leq 3h_{j-1}-qh_{j-1}=(3-q)h_{j-1}.$$
        \end{itemize}
        \end{itemize}
        However, we cannot simply charge this excess to $B$.  We need to show that the requests causing positive excess were already known when the previous phase decided to continue.
        Therefore, we now prove the following: Only the requests known at the previous checkpoint $tc=(1+\alpha)h_{j-1}$ can cause positive excess.
        Suppose the preceding phase used the scheduled checkpoint $tc=(1+\alpha)h_{j-1}$. 
        Suppose a request $r$ arrives after $tc$: $a_r>(1+\alpha)h_{j-1}.$
        $r$'s service time in the next phase must satisfy $c'\geq a_r>(1+\alpha)h_{j-1}.$
        Since $q=1+2\alpha$ and $\alpha\in (1/2,1)$, $q(1+\alpha)>3.$ Therefore, $\mathcal{E}(c')=3h_{j-1}-qc'<3h_{j-1}-q(1+\alpha)h_{j-1}<0.$ Hence, a request arriving after the previous checkpoint cannot produce positive excess, which concludes
        \begin{align}
        \mathcal{E}(c')>0\implies a_r\leq (1+\alpha)h_{j-1}.
        \end{align}

        Now let us connect this to the set $\mathcal{Q}_{j-1}$. Recall
        \begin{align}
        \mathcal{Q}_{j-1}=\{r:a_r\leq (1+\alpha)h_{j-1}, r\notin \mathcal{D}\cup \mathcal{S}_{j-1}\}.
        \end{align}
        Suppose $r\in \mathcal{F}_j$ has positive excess. By Equation 5, $r$ was already released at the previous checkpoint.
        There are two possibilities:
        \begin{itemize}
            \item If $r$ was in $\mathcal{D}$, it was already registered and cannot be first-registered in phase $j$.
            \item If $r$ was in $\mathcal{S}_{j-1}\setminus \mathcal{D}$, the previous phase was complete and therefore registered every request in that set.
        \end{itemize}
        Thus, $r$ must have been outside $\mathcal{D}\cup \mathcal{S}_{j-1}.$ Consequently, 
        \begin{align}
            \{r\in \mathcal{F}_j:\mathcal{E}(c')>0\}\subseteq \mathcal{Q}_{j-1}.
        \end{align}
        Therefore, Equation 6 tells that the excess in the next phase comes only from requests already known when the previous phase made its decision. 

        We now bound the total excess of the next phase. We know that every request has excess at most $(3-q)h_{j-1}$. From Equation 7, only requests in $\mathcal{Q}_{j-1}$ can have positive excess. 
        Therefore, 
        \begin{align}
        \sum_{r\in \mathcal{F}_j}w_r\max\{\mathcal{E}(c'),0)\}\leq (3-q)h_{j-1}w(\mathcal{Q}_j-1).
        \end{align}

        The question is where do we get enough allowance to pay for this excess. We now show that the tail of the previous phase provides the allowance. Consider the completed phase $j-1$ with start time $h_{j-1}$. Recall its tail $$\mathcal{T}_{j-1}=\{r\in \mathcal{A}_{j-1}:c_{(j-1)r}>\alpha h_{j-1}\}.$$
        Because phase $j-1$ was completed, it executed $P_{j-1}$ until elapsed time $h_{j-1}$.
        Every request in $\mathcal{A}_{j-1}$ has virtual service time strictly below $h_{j-1}$. Therefore, every such request was first-registered: $\mathcal{F}_{j-1}=\mathcal{A}_{j-1}$. In particular, $\mathcal{T}_{j-1}\subseteq \mathcal{F}_{j-1}$. These requests have already earned their contributions to $B$.
        For $r\in \mathcal{T}_{j-1},$ $c_{(j-1)r}>\alpha h_{j-1}.$ We also know that $h_{j-1}\geq qh_{j-2}$. Since $q\alpha>1$ for $\alpha\in (1/2,1)$, it follows that $\alpha h_{j-1}>\frac{h_{j-1}}{q}\geq h_{j-2}.$ Thus, $\beta_{(j-1)r}=c_{(j-1)r}.$
        The completion time upper bound for this request is $H_{(j-1)r}=h_{j-1}+c_{(j-1)r}.$
        Its unused allowance is therefore
        \begin{align}
        K\beta_{(j-1)r}-H_{(j-1)r} &= Kc_{(j-1)r}-(h_{j-1}+c_{(j-1)r})\\
        &=(K-1)c_{(j-1)r}-h_{j-1}\\
        &=qc_{(j-1)r}-h_{j-1}.
        \end{align}
        Using $c_{(j-1)r}>\alpha h_{j-1}$,
        \begin{align}
          K\beta_{(j-1)r}-H_{(j-1)r} &=(q\alpha-1)h_{j-1}.  
        \end{align}
        Summing over the tail,
        \begin{align}
          \sum_{r\in\mathcal{T}_{j-1}}w_r\left(K\beta_{(j-1)r}-H_{(j-1)r}\right) >(q\alpha-1)h_{j-1}w(\mathcal{T}_{j-1}), 
        \end{align}
        which is the unused allowance supplied by the late tail. 

        We now show why this allowance is sufficient. Algorithm \ref{algorithm:capped} chooses to finish the phase $j-1$ only if $w(\mathcal{Q}_{j-1})<\theta w(\mathcal{T}_{j-1}).$
        Multiplying the above equation by $(3-q)h_{j-1}$, we obtain
        \begin{align}
        (3-q)h_{j-1}w(\mathcal{Q}_{j-1})<(3-q)h_{j-1}\theta w(\mathcal{T}_{j-1}).
        \end{align}
        Substituting $\theta=\frac{q\alpha-1}{3-q}$, we obtain
        \begin{align}
        (3-q)h_{j-1}w(\mathcal{Q}_{j-1})<(q\alpha-1)h_{j-1} w(\mathcal{T}_{j-1}).
        \end{align}
Comparing Equations 8, 13, and 15, we obtain
\begin{align}
        \sum_{r\in \mathcal{F}_j}w_r\max\{\mathcal{E}(c'),0)\}\leq \sum_{r\in\mathcal{T}_{j-1}}w_r\left(K\beta_{(j-1)r}-H_{(j-1)r}\right).
        \end{align}

We now sum allowances across phases. We combine two cases.
Let the total positive excess be
$$X=\sum_j\sum_{r\in \mathcal{F}_j} \max\{w_r(H_{jr}-K\beta_{jr}),0\}.$$ 
Let the total positive allowance be
$$Y=\sum_j\sum_{r\in \mathcal{F}_j} \max\{w_r(K\beta_{jr}-H_{jr}),0\}.$$ 

We established that (i) the first phase and all phases following cuts create no positive excess, and (ii) the positive excess following each long phase is paid for by that phase's tail allowance (Equation 16). Furthermore, the sets $\mathcal{T}_j$ belonging to different long phases are disjoint. This is because $\mathcal{T}_j\subseteq\mathcal{F}_j$, and every request is first-registered in exactly one phase. Therefore, no request can belong to two different first-registered sets $\mathcal{F}_j$. Also, each long phase's tail pays only for the immediately following phase. Therefore, no allowance is spent twice, giving $X\leq Y.$

Now observe that, for every real number $z$, $z=\max\{z,0\}-\max\{-z,0\}$. Consequently,  
$$\sum_j\sum_{r\in \mathcal{F}_j}w_r(H_{jr}-K\beta_{jr})=X-Y\leq 0.$$

Hence,
$$\sum_j\sum_{r\in \mathcal{F}_j}w_rH_{jr}\leq K \sum_j\sum_{r\in \mathcal{F}_j} w_r\beta_{jr}.$$

Recalling the definitions of $H_{jr}$ and $B$,
$$\sum_j\sum_{r\in \mathcal{F}_j}w_r(h_j+c_{jr})\leq KB.$$
Therefore, $\mathrm{ALG}\leq KB$ and the lemma follows.
\end{proof}

\vspace{2mm}
\noindent{\bf Proof of Theorem 1 ($1$-TRP).}
\begin{proof}
We have $\alpha\in (1/2,1)$. We have set $q=1+2\alpha$, 
$K=1+q=2+2\alpha$, $\theta=\frac{(2\alpha-1)(1+\alpha)}{2(1-\alpha)}$, and $M=\frac{\theta}{1-\alpha}=\frac{(2\alpha-1)(1+\alpha)}{2(1-\alpha)^2}$ in Algorithm \ref{algorithm:capped}.  We have that $q\alpha>1.$
From Lemma \ref{lemma:BOPT}, we have that
$B\leq \left(1+\frac{q}{M(q-1)}\right)OPT$ and from Lemma \ref{lemma:KB}, $\mathrm{ALG}\leq KB.$
The resulting ratio becomes
$$\frac{\mathrm{ALG}}{\mathrm{OPT}}\leq K\left(1+\frac{q}{M(q-1)}\right)=\frac{2(4\alpha^3-2\alpha^2-\alpha+1)}{\alpha(2\alpha-1)}.$$
The unique minimizer for $\frac{\mathrm{ALG}}{\mathrm{OPT}}$ with $\alpha\in (1/2,1)$ is $\alpha^*\approx0.892$. Therefore,
$\min_{\alpha} \frac{\mathrm{ALG}}{\mathrm{OPT}}\approx 3.877.$
\end{proof}

If we set $\alpha=1$ in Algorithm \ref{algorithm:capped}, the phase-$j$ horizon grows by exactly 3 times the phase-$(j-1)$ horizon, i.e.,  $h_j=3h_{j-1}$, irrespective of the cut. Therefore, we have the following corollary.
\begin{corollary}
    Setting $\alpha=1$, Algorithm \ref{algorithm:capped} achieves 4-competitive ratio of \cite{BienkowskiKL21-ICALP}. 
\end{corollary}

\subsection{Analysis of the Algorithm -- Zero-Residual Instances}
We now prove Theorem \ref{theorem:main} 
considering the special class of zero-residual instances. 
In Algorithm \ref{algorithm:capped}, the residual cost is denoted by $\sum_j e_j$, which is the capped-route cost loss due to interruptions.

\begin{definition} [Zero-residual Execution] 
\label{def:zero-residual-execution}
An execution of Algorithm \ref{algorithm:capped} is {\em zero-residual} if 
$e_j=0$ for every phase $j$. 
\end{definition}
This means that whenever the capped route $P_j$ selects a request in phase $j$, that request is registered before phase $j$ ends. 
In other words,
the newly arriving requests during phase $j$ never make $P_j$ sufficiently unattractive to trigger an interruption.
Let the last phase of Algorithm \ref{algorithm:capped} be called  {\em terminal phase}. All previous phases are non-terminal phases. A {\em strictly interruption-free} execution is a special case of the zero-residual execution: its non-terminal phases execute the complete capped route (no interruptions cutting the route). 
The zero-residual execution is more general than the strictly interruption-free execution since it permits the server stop traversing $P_j$ at phase $j$ and return to $o$ whenever the tail become empty before cap $h_j$, i.e., $\mathcal{T}_j=\emptyset$. 

\begin{definition} [Zero-residual Instance] 
\label{def:zero-residual-instance}
An instance that satisfies Definition \ref{def:zero-residual-execution} in Algorithm \ref{algorithm:capped}. 
\end{definition}

\vspace{2mm}
\noindent{\bf Proof of Theorem 1 ($1$-TRP, Zero-residual Instances).}
\begin{proof}
We prove the deterministic $(3+\epsilon)$-competitive guarantee for Algorithm \ref{algorithm:capped}, for every $\epsilon>0$. 
Since no non-terminal phase is interrupted, every phase $j$ executes its $P_j$ completely until cap $h_j$: $u_j=h_j$. Consequently, every request selected by the optimal capped oracle is registered in that phase:
$\mathcal{F}_j=\mathcal{A}_j$. Therefore, 
for correctness, Lemma \ref{lemma:serve} extends as it already works for both no cut and cut situations, i.e., every request is eventually registered on a finite set $\mathcal{R}$
 of requests.

 We now prove the competitive ratio.
Recall from Lemma \ref{lemma:ej}
$$e_j=\sum_{r\in \mathcal{A}_j\setminus \mathcal{F}_j} w_r(h_j-c_{jr}).$$
Since $\mathcal{F}_j=\mathcal{A}_j$, $e_j=0$.
Recall from Lemmas \ref{lemma:B} and \ref{lemma:BOPT}
$B\leq OPT+\sum_je_j$,
where   $B=\sum_j\sum_{r\in \mathcal{F}_j}w_r\max\{h_{j-1},c_{jr}\}.$
Since every $e_j=0$ in a zero-residual instance, we immediately obtain $B\leq \mathrm{OPT}$. 

We now prove that $\mathrm{ALG}$ is still bounded with respect to the same terms $K$ and $B$: $\mathrm{ALG}\leq KB$. This is the subtle part. Notice that, since every phase is completed, the phase horizon grows by a factor of $3$: $h_{j+1}=3h_j$. Therefore, we cannot simply claim that every request $r$ satisfies 
$C_r\leq K \max\{h_{j-1},c_{jr}\}.$ That is, some requests registered immediately after a completed phase have a larger physical-cost upper bound. 

We prove that the original long-phase charging argument we used in Lemma \ref{lemma:KB} still applies even when there is no cut. 
Let $K=1+q=2+2\alpha, \alpha\in (1/2,1)$ as before. For requests registered in phase $1$ or following an interrupted phase, the physical-cost inequality holds directly. For requests registered after a completed phase,
we show that the excess cost is bounded by
$X_j\leq (3-q)h_jw(\mathcal{Q}_j),$ where $\mathcal{Q}_j$ consists of requests omitted by phase $j$ but released by its checkpoint.
Let $j$ be the first registration phase of request $r$. Its virtual registration cost $\beta_{jr}=\max\{h_{j-1},c{jr}\}$ and its physical-cost upper bound is $H_{jr}=h_j+c_{jr}.$
Since we consider zero-residual instances, phase $j-1$ must be complete: $h_j=3h_{j-1}.$
The excess of the physical-cost upper bound over its $K$-allowance is 

$$E(r)=3h_{j-1}+c_{jr}-K\max\{h_{j-1},c_{jr}\}.$$ We first show that the positive excess can only occur for the requests included in $\mathcal{Q}_j$, where
$\mathcal{Q}_j=\{r:a_r\leq (1+\alpha)h_{j-1}\}\setminus(\mathcal{D}_{j-1}\cup \mathcal{S}_j).$ If $c_{jr}>(1+\alpha)h_{j-1}$, then $c_{jr}\geq a_r>(1+\alpha)h_{j-1}.$ Also, $q(1+\alpha)>3$ for $\alpha>1/2$. Therefore, $c_{jr}\geq h_{j-1}$ and $E_r\leq (3-q)h_{j-1}.$ Therefore, following the analysis as in Lemma \ref{lemma:KB} for $c_{jr}\leq h_{j-1}$ and $c_{jr}>h_{j-1}$, we have that the total positive excess of requests first-registered in phase $j+1$ is at most 
$X_j\leq (3-q)h_{j-1}w(\mathcal{Q}_j$. 

The completed phase condition gives $w(\mathcal{Q}_j)<\theta w(\mathcal{T}_j).$ The unused physical-cost allowance of the tail requests $\mathcal{T}_j$ is at least $(q\alpha-1)h_jw(\mathcal{T}_j).$ Using $(3-q)\theta=q\alpha-1,$ following the analysis in Lemma \ref{lemma:KB}, the unused allowance pays for all positive physical-cost excess in phase $j+1$. 
Finally, the sets $\mathcal{T}_j$ are disjoint across completed phases because every member of $\mathcal{T}_j$ is first-registered in phase $j$. Hence no unused allowance is spent twice.
Therefore, even when all phases are completed,
$\mathrm{ALG}\leq KB.$ Substituting $B$ by $\mathrm{OPT},$ we obtain
$$\mathrm{ALG}\leq K \cdot \mathrm{OPT}=(2+2\alpha)\mathrm{OPT},$$ for every $\alpha\in (1/2,1)$.
Setting $\alpha=1/2+\epsilon, \epsilon>0,$ gives $\frac{\mathrm{ALG}}{\mathrm{OPT}}=3+\epsilon$. Setting $\alpha=1$, the ratio matches the 4 competitive guarantee of \cite{BienkowskiKL21-ICALP}.  
\end{proof}

\begin{algorithm}[bt!]
{
\footnotesize
All $k$ servers begin at the depot. Let $m=\min_r\max\{a_r,d(o,x_r)\}$, and initialize the first phase horizon to $h_1=qm$

At the beginning of phase $j$, compute the optimal $k$-server capped routes  $\{P_{j,1},P_{j,2},\ldots,P_{j,k}\}$ with cap $h_j$

Start executing all $k$ virtual routes simultaneously at absolute time $h_j$

At every distinct time $t, h_j<t<tc$, conduct an emergency test using
$
\mathcal{N}(t)$ and 
$\mathcal{V}_j(u)$, $u=t-h_j$, as in Algorithm \ref{algorithm:capped} 

At the scheduled checkpoint $tc=(1+\alpha) h_j$, apply the same checkpoint test using $\mathcal{T}_j$ and $\mathcal{Q}_j$ as in Algorithm \ref{algorithm:capped} 

In case of emergency or scheduled cut at phase $j$, all servers return to the depot and start  phase $j+1$ at time $qh_j$

If phase $j$ is completed, all servers return to the depot and start phase $j+1$ at time $3h_j$

All servers stop at $o$ if, after a phase, they find that there is no request remains to be served

     \caption{Modified Algorithm for $k$-TRP, $k\geq 2$}
\label{algorithm:modified-capped}
}
\end{algorithm}

\section{$k$-TRP Algorithm, $k\geq 2$}
\label{sec:k-TRP}
We now discuss how Algorithm \ref{algorithm:capped} 
extends to $k$-TRP. The modified algorithm is given in Algorithm \ref{algorithm:modified-capped}. All $k$ servers are initially at the depot $o$. 
At phase $j$, let $\mathcal{R}(h_j)$ be the requests revealed so far. The optimal capped-route oracle chooses $k$ optimal routes 
$\mathcal{P}_j=\{P_{j,1},P_{j,2},\ldots,P_{j,k}\}$ all starting at $o$ with cap $h_j$. Define $c_{jr}=\min_{1\leq s\leq k}c_{jr}^s,$ where $c_{jr}^s$ is the earliest feasible service time for request $r$ on server $s$'s route $P_{j,s}$. 
If no route serves $r$, set $c_{jr}=\infty$.
The oracle solves
$$\mathcal{L}_k(h_j)=\min_{P_j}\sum_{r\in \mathcal{R}_j}w_r\min\{c_{jr},h_j\}.$$
All $k$ routes are executed simultaneously at horizon $h_j$. Each request is assigned to a server that attains its earliest virtual service time, with deterministic tie-breaking. The capped objective counts each request only once, even though there are $k$ routes. We have the following proof.

\vspace{2mm}
\noindent{\bf Proof of Theorem 1 ($k$-TRP, $k\geq 2$).}

\begin{proof}
Suppose a server $s$ starts at the depot and follows its route $P_{j,s}$ in phase $j$ for $u$ time units ($u$ is the elapsed time since $h_j$). Its position at that time, say $x$, satisfies $d(o,x)\leq u$, because its speed is at most one. Consequently, it can return to $o$ in  another $\leq u$ time units.
If the route $P_{j,s}$ of server $s$ is cut in phase $j$ at elapsed time $u$, all $k$ servers return to $o$ simultaneously.   
Since a cut in phase $j$ happens at elapsed time $u\leq \alpha h_j$,
a cut in phase $j$ starts phase $j+1$ at horizon $h_{j+1}=qh_j=(1+2\alpha)h_j, \alpha\in (1/2,1)$ as in Algorithm \ref{algorithm:capped}. No cut in phase $j$ starts phase $j+1$ at $h_{j+1}=3h_j$. 

Let $\mathrm{OPT}_k$ denote the optimal offline cost with $k$ servers.  
We now argue that the following inequality holds: $B\leq \mathrm{OPT}_k+\sum_je_j,$
where (as in Algorithm \ref{algorithm:capped}) $\Delta_j=\mathcal{L}_k(h_j)-\mathcal{L}_k(h_{j-1})$, $e_j=\sum_{r\in \mathcal{A}_j\setminus\mathcal{F}_j}w_r(h_j-c_{jr}),$ and $B=\sum_j\sum_{r\in \mathcal{F}_j}w_r\max\{h_{j-1},c_{jr}\}.$

We fix the optimal $k$-server capped-route collection $\mathcal{P}_j$ chosen by the oracle at cap $h_j$. We then evaluate this same collection of routes at the cap $h_{j-1}$ of phase $j-1$. $\mathcal{P}_j$  is a feasible candidate for the smaller-cap oracle. Therefore,
$\mathcal{L}_k(h_{j-1})
\le
\sum_{r\in \mathcal{R}_j}w_r\min\{c_{jr},h_{j-1}\}.$
Subtracting this from the objective $\mathcal{L}_k(h_{j})$ at cap $h_j$ gives
$$\Delta_j\ge
\sum_{r\in \mathcal{R}_j}w_r
\left(
\min\{c_{jr},h_j\}
-\min\{c_{jr},h_{j-1}\}
\right).$$
In other words, $\Delta_j$ depends on the optimality of the capped-route oracle, not on the number of routes.
The same request-by-request accounting then gives
$B\le \mathcal{L}_k(h_j)+
\sum_j(\Delta_j+e_j),$
with the initial credit $\mathcal{L}_k(h_1)$ handled as in the original initialization argument. Since the $k$-server optimal offline algorithm $\mathrm{OPT}_k$ is a feasible candidate for the capped objective,
$\mathcal{L}_k(h_j)\le \mathrm{OPT}_k$ as in the $k=1$ case in Lemma \ref{lemma:LhjOPT}. 
As in Lemma \ref{lemma:serve}, every request is eventually first-registered since the same argument applies for each of the $k$ routes. 
Therefore, the $\Delta_j$ values telescope, yielding $B\leq \mathrm{OPT}_k+\sum_je_j$ as in Lemma \ref{lemma:BOPT}. 

The capped-route cost loss (residual cost) $e_j$ is bounded in the same way as in Lemma \ref{lemma:ejOPT} giving $\sum_je_j=\frac{q}{M(q-1)}\mathrm{OPT}_k$.
Plugging this in $B$ gives $$B\leq \mathrm{OPT}_k+\sum_je_j\leq  (1+ \frac{q}{M(q-1)})\mathrm{OPT}_k,$$ with $M$ and $q$ parameters as in Algorithm \ref{algorithm:capped}.  The physical-cost bound $\mathrm{ALG}_k$ also extends since each of the routes $P_{j,s}$ is each phase $j$ are optimal. Suppose request $r$ is first-registered in phase $j$. It is first-registered at time $h_j+c_{jr}$, on one of the $k$ routes.   Therefore, $\mathrm{ALG}_k\leq KB.$ Setting the values of the parameters $q,K,\theta,$ and $M$ as in Algorithm \ref{algorithm:capped} with the unique minimizer $\alpha^*\approx0.892$ when $\alpha\in (1/2,1)$, we obtain $\min_{\alpha} \frac{\mathrm{ALG}_k}{\mathrm{OPT}_k}\approx  3.877.$ If we set $\alpha=1$, we match the 4 ratio of \cite{BienkowskiKL21-ICALP}. 

The analysis for the zero-residual instance extends combining the 1-TRP zero-residual analysis with the above $k$-TRP analysis, giving the same  $\frac{\mathrm{ALG}_k}{\mathrm{OPT}_k}=3+\epsilon$ ratio for every $\epsilon>0$.
\end{proof}

\section{$k$-DARP, $k\geq 1$}
\label{sec:k-DARP}
We now discuss how the $k$-TRP results of Sections \ref{sec:1-TRP} and \ref{sec:k-TRP} extend to $k$-DARP, $k\geq 1$. 
The lower bound for $k$-DARP (any variant) is 2  with $k\geq 2$ \cite{bonifaci2009-TCS}. 
For uncapacitated $1$-DARP, the lower bound is $1+\sqrt{2}\approx 2.414$ \cite{feuerstein2001line}.
For capacitated 1-DARP with server capacity 1,  the lower bound is $3$ \cite{feuerstein2001line}.   The state-of-the-art upper bound  is 4 (like $k$-TRP) for any variant of  $k$-DARP \cite{BienkowskiKL21-ICALP}. 

We establish two results: 
\begin{itemize}
    \item {\bf Uncapacitated $k$-DARP, $k\geq 1$:} Deterministic 3.877-competitive guarantee ({\bf Section \ref{subsec:uncapacitated}}). 
    \item {\bf Capacitated, Non-Preemptive $k$-DARP, $k\geq 1$:} Deterministic $(3+\epsilon)$-competitive guarantee, $\epsilon>0$,  for zero-residual instances ({\bf Section \ref{subsec:capacitated}}).
\end{itemize}

\noindent{\bf DARP Specific Notation.}
In DARP, each request $r$ is specified by a quadruple $r=(a_r,s_r,d_r,w_r)$, where $a_r\geq 0$ is its release time, $s_r\in \mathbb{R}^2$ is its pickup location, $d_r\in \mathbb{R}^2$ is its delivery location, and $w_r>0$ is its weight. 
TRP is the special case of DARP in which each request has identical pickup and delivery locations and hence can be written simply as $x_r$ to denote both $s_r$ and $d_r$. We assume (at most) unit-speed $k \geq 1$ servers initially at the depot $o$, instantaneous pickup and delivery, and no maximum ride-time restriction.  
A request is {\em complete} only when its passenger reaches $d_r$. The objective is 
$\mathrm{ALG}=w_rC_r,$ where $C_r$ is the passenger's delivery time. 

\subsection{Uncapacitated $k$-DARP}
\label{subsec:uncapacitated}

In uncapacitated $k$-DARP, each server has infinite capacity.  
Preemption becomes irrelevant with uncapacitated servers as the passengers may remain onboard during returns to the depot and while waiting there. 
The main change w.r.t. TRP algorithms is that a route $P$ now contains two events for each request $r$:
$\mathrm{pickup}(r)$: $s_r$ and $\mathrm{delivery}(r)$: $d_r$. Therefore, the oracle must ensure $p_{Pr}\geq a_r$ and $c_{Pr}\geq p_{Pr}$, where $p_{Pr}$ and $c_{Pr}$ denote the virtual pickup and delivery times in route $P$. 

Define 
$$\mathcal{L}_D(h_j)=\min_{P}\sum_rw_r\min\{c_{Pr},h_j\},$$
with $P$ ranging over feasible pickup-delivery routes starting at the depot. $\mathcal{L}_D(h_j)$ optimizes $P$ over precedence-respecting orders of pickup and delivery events based on the set $\mathcal{R}(h_j)$ of requests known until time $h_j$. For requests not yet released at time $h_j$, their capped contribution is simply $w_rh_j$, because their delivery cannot occur before their release. Thus, we can optimize $P_j$ using only the requests in  $\mathcal{R}(h_j)$. The important change in DARP compared to TRP is the cost is now based on delivery time, not pickup time. Therefore, define the earliest possible delivery lower bound for request $r$ by
$$b_r=\max\{a_r,d(o,s_r)\}+d(s_r,d_r).$$
This accounts for the server's need to reach the pickup location after release and then travel to the delivery location. Set $m=\min_rb_r.$ Every offline delivery time satisfies $C^*_r\geq b_r\geq m.$ Thus,
$\mathcal{L}_D(m)=m\sum_rw_r$. As the phase horizons grow geometrically, eventually every request has virtual delivery time below the phase cutoff. Therefore, Lemma \ref{lemma:serve} applies. 
The key inequality is still
$$(h_j-h_{j-1})w(\mathcal{Q})\leq \mathcal{L}_D(h_j)-\mathcal{L}_D(h_{j-1}).$$ 

For an emergency interruption, new requests have release times after $h_j$, so their virtual delivery times are also after $h_j$. They therefore contribute the full increment $h_j-h_{j-1}$.
For a scheduled interruption, a request with $c_{jr}>\alpha h_j$ loses at most
$(1-\alpha)h_jw_r.$
We need the following lemma so that we have the precise argument for how we deal with the requests that have not been delivered due to the cut. 
%
Partition $R(h_j)$ into: $R(h_j)=\mathcal{O}(h_j)\cup \mathcal{N}(h_j), \mathcal{O}(h_j) \cap \mathcal{N}(h_j)=\emptyset,$ where $\mathcal{O}(h_j)$ denotes requests whose passengers are already onboard and $\mathcal{N}(h_j)$ denotes requests whose passengers have not yet been picked up. Consider the route $P_j$. $P_j$ contains the requests in $\mathcal{R}(h_j)$ including $\mathcal{O}(h_j).$

\begin{lemma}
\label{lemma:ghost}
Suppose,  for each request $r$, its status is one of: (i) not yet picked up;
(ii) already onboard;
(iii) already delivered.
Let $P_j$ be a feasible virtual pickup-delivery route for every request in $\mathcal{R}(h_j)$ (released by time $h_j$).
Execute $P_j$ physically as follows:
\begin{itemize}
    \item For $r\in \mathcal{N}(h_j)$ request, execute its pickup event  $s_r$.
\item For $r\in \mathcal{O}(h_j)$, treat $s_r$ as a {\em ghost pickup}: the server follows $P_j$  through $s_r$, but does not physically pick up the passenger again. 
\item For an already delivered request, skip both physical service operations, but retain the virtual route's movement.
\item Execute the delivery event $d_r$ normally for every passenger that has not been already delivered.
\item At interruption, return to the depot and keep all undelivered passengers onboard.
\end{itemize}
Then the physical execution is feasible and never picks up a passenger twice.
Moreover, if request $r$ is not already delivered, its physical delivery time satisfies
$C_r\le h_j+c_{jr}$.
\end{lemma}

\begin{proof}
Initially, all passengers in the onboard set are physically onboard. Every unpicked passenger is absent from the vehicle, and every delivered passenger is no longer present.

We induct over the events of the capped route $P_j$.

At a pickup event for request $r$:
\begin{itemize}
    \item  If $r\in \mathcal{O}(h_j)$ (already onboard), the pickup is ghosted. No physical operation occurs.
\item  If $r\in \mathcal{N}(h_j)$ (not yet been picked up), the physical pickup is executed.
\item  If $r$ is already delivered, the physical pickup is skipped.
\end{itemize}
Thus, no passenger is picked up twice.

Now consider a delivery event for request $r$. Since the virtual route satisfies pickup-before-delivery, there are two relevant possibilities:
\begin{itemize}
\item The passenger was already onboard at the beginning of the phase, i.e., $r\in \mathcal{O}(h_j)$. The virtual pickup is ghosted, so the passenger remains onboard until the delivery event.
\item  The passenger was not onboard initially and has not already been delivered. Its virtual pickup event was physically executed earlier, so it is onboard at delivery  (i.e., $r\in \mathcal{O}(h_j)$).
\end{itemize}
Hence every physical delivery is feasible.

For the time bound, execute the capped route $P_j$ at horizon $h_j$ with cap $h_j$. Every request included in $P_j$ was released by $h_j$, so the schedule respects its release time. Ghosting a pickup or delivery does not change the trajectory of $P_j$. Thus, if the virtual delivery event occurs at time $c_{ji}$ in $P_j$, the physical delivery occurs no later than real time $h_j+c_{ji}$ traversing $P_j$.

At interruption, passengers who have not been delivered remain onboard during the return. Infinite capacity ensures that retaining them does not restrict the route.
\end{proof}


Since the the capped route oracle measures delivery rather than first visit to a location, the same capped-cost increment $\Delta_j$ comparison gives $B\leq \mathrm{OPT}+\sum_je_j.$ For a finite set $\mathcal{R}$ of requests with $w_r>0$, sufficiently large phase horizons force all requests to be deliverable within the cap. The proof of $e_j$ in Lemma \ref{lemma:ejdeltaj} extends as it uses only the capped-cost objective, request weights, and the phase thresholds and does not depend on wether a request is a point request like TRP or a pick-up delivery pair. With Lemma \ref{lemma:ghost}, the same phase-to-phase charging argument gives $\mathrm{ALG}\leq KB$, as additional cost incurred by carrying passengers between phases is ignored. Setting parameters as we used before in the proof of Theorem \ref{theorem:main}, we obtain the 3.877 ratio. For $k$-DARP, the ideas extend modifying everything respecting the availability of $k$ servers as in the $k$-TRP case (Section \ref{sec:k-TRP}).
Therefore, we summarize the main result in the following theorem.

\begin{theorem}
    For uncapacitated $k$-DARP, $k\geq 1$, ignoring additional cost caused by carrying passengers between phases, $\frac{\mathrm{ALG}_k}{\mathrm{OPT}_k}\approx 3.877$.
\end{theorem}

\subsection{Capacitated, Non-preemptive $k$-DARP -- Zero-Residual Instances}
\label{subsec:capacitated}
We assume that each server has capacity $C\geq 1$. We do not allow preemption. We consider zero-residual instances (Definition \ref{def:zero-residual-instance}) meaning that interruptions (emergency or scheduled) are not triggered. 
In the algorithm for capacitated, non-preemptive $k$-DARP, we need the 
 following guarantee as in \cite{BienkowskiKL21-ICALP}: every passenger is either picked up and delivered within the phase, or is not picked up at all.
We guarantee this through atomic routes.
\begin{definition}[Atomic Feasible Route]
\label{def:atomicfeasibleroute}
    For a phase horizon $h_j$, a capped route $P_j$ is called {\em atomic feasible route} if 
    \begin{itemize}
        \item Every request $r$ served by $P_j$ is picked up after  time $a_r$;
\item Every served request $r$ is delivered after its pickup;
\item At every point, the number of onboard passengers is at most $C$;
\item Every passenger picked up by $P_j$ is delivered by virtual time $h_j$ (absolute time $2h_j$).
    \end{itemize}
\end{definition}
Definition \ref{def:atomicfeasibleroute} guarantees that all other requests are left untouched as in \cite{BienkowskiKL21-ICALP}. The optimal route oracle becomes
$$\mathcal{L}(h_j)=\min_{P~ \text{atomic and feasible}}\sum_rw_r\min\{c_{Pr},h_j\}.$$ The minimization is over the requests revealed by time $h_j$. The request arrive at any time $t>h_j$ contribute the penalty $h_jw_r$, so they do not affect the route choice. 
With the capacity $C$, it is impossible for a route to pick up a passenger and leave the passenger onboard at the phase boundary.
We can establish the following lemma for a capped route satisfying Definition \ref{def:atomicfeasibleroute}.

\begin{lemma}
\label{lemma:truncation}
    Let $P_{j}$ be an atomic feasible schedule of horizon $h_j$. For every $0<h_p<h_j$, there is an atomic feasible schedule $P_{p}$ of horizon $h_p$ satisfying 
    $$\min\{c_{pr},h_p\}=min\{c_{jr},h_p\}, \text{~for every request}~r.$$
\end{lemma}
\begin{proof}
Keep the same trajectory as $P_j$, but truncate it at time $h_p$.
For each request $r$: (i) if its delivery occurs by time $h_p$, retain its pickup and delivery actions, and (ii) if its delivery occurs after time $h_p$, remove both its pickup and delivery actions. 
Because $P_j$ is atomic, every passenger picked up by time $h_p$ but not delivered by time $h_p$ can be omitted entirely. Removing these passengers cannot increase the number of passengers onboard at any time.

The retained pickup-delivery pairs preserve release times and precedence. The trajectory $P_j$ is unchanged, so the retained delivery times are unchanged.
Consequently, the new schedule is feasible, respects capacity, and satisfies $\min\{c_{pr},h_p\}=min\{c_{jr},h_p\},$ for every request $r$.
\end{proof}

Let $P_j$ be an optimal atomic capped route for horizon $h_j$. 
By Lemma \ref{lemma:truncation}, its truncation is feasible for horizon $h_{j-1}$. Therefore,
$$\mathcal{L}(h_j)-\mathcal{L}(h_{j-1}\geq \sum_rw_i(\min\{c_{jr},h_j\}-\min\{c_{(j-1)r},h_{j-1}\}).$$

Now consider an optimal offline DARP route and truncate it at time $h_j$, discarding both actions of every passenger not delivered by then. This remains capacity-feasible. Restricting it to the requests $\mathcal{R}(h_j)$ revealed until $h_j$ cannot make it infeasible. Thus,
$$\mathcal{L}(h_j)\le \sum_r w_r\min\{C^*_r,h_j\}\le \mathrm{OPT}.$$

Here $C^*_r$ is the optimal absolute delivery time of request $r$. The request-wise accounting we used in our proofs for $B$ applies as the capped-cost increments pay for the first-registration credits, while $e_j$ accounts for selected requests that are not registered in the phase. Therefore, 
$B\leq \mathrm{OPT}+\sum_j e_j$. If every phase is completed without interruption, then $\mathcal{F}_j=\mathcal{A}_j$, so $e_j=0$. Therefore, $B\leq \mathrm{OPT}.$
Due to Lemma \ref{lemma:truncation}, when phase $j$ starts at time $h_j$, its optimal capped route $P_j$ is traversed completely: every passenger picked up during this traversal is delivered before the traversal of $P_j$ finishes. The server therefore returns to the depot $o$ empty and hence the capacity-feasible route  is guaranteed by the oracle. Therefore, following our $\mathrm{ALG}$ cost analysis for $k$-TRP, we obtain $\mathrm{ALG}\leq KB$ since the atomic capped-route oracle uses the same phase and checkpoint rules. 
We summarize the main result in the following theorem.

\begin{theorem}
\label{theorem:capacitated-non-preemptive-DARP}
    For zero-residual instances of non-preemptive $k$-DARP with each server having capacity $C\geq 1$, $\frac{\mathrm{ALG}}{\mathrm{OPT}}\leq 2+2\alpha$ for every $\alpha\in (1/2,1).$ Consequently, $\frac{\mathrm{ALG}}{\mathrm{OPT}}=3+\epsilon$, setting $\alpha=1/2+\epsilon, \epsilon>0.$
\end{theorem}

We make the following observation for the capacitated $1$-DARP with server capacity $C=1$. 

\begin{observation}
\label{obs:tight}
    For zero-residual instances of $1$-DARP with server capacity $C=1$, Theorem \ref{theorem:capacitated-non-preemptive-DARP} is (almost) tight given the 3 lower bound of \cite{feuerstein2001line}. 
\end{observation}

\section{Discussion and Future Directions}
\label{sec:conclusion}
The proposed interruptible capped-route framework  has improved the deterministic competitive guarantee for online $k$-TRP, $k\geq 1$,  from the state-of-the-art 4 to 3.877 by combining the phase-based optimal capped-route oracle with two kinds of adaptive interruptible decisions (emergency and scheduled) within each phase. Additionally, this framework allowed us to  analyze the special class of zero-residual instances for which our framework does not trigger an interruption, giving a substantially improved $(3+\epsilon)$-competitive guarantee, for every $\epsilon>0$. These results extend to some variants of online $k$-DARP. For capacitated 1-DARP with server capacity 1, our bound for the zero-residual instances is (almost) tight (Observation \ref{obs:tight}), the first such result to (almost) close the gap between the upper and lower bounds. We believe our results open the door for future research in  
%
%
%
several questions we list, among many others. 

\vspace{2mm}
\noindent{\bf Time Complexity of the Capped-Route Oracle.}
As in \cite{BienkowskiKL21-ICALP},
for a finite set of revealed requests, computing an optimal capped route at every phase involves enumerating subsets and service orders and evaluating their earliest feasible service times. 
This provides a finite deterministic implementation of the achievable competitive ratio, but the number of candidate orders can be exponential in the number of requests and hence becomes equivalent to solving an $NP$-hard task. Accordingly, like  \cite{BienkowskiKL21-ICALP}, our competitive analysis does not establish a polynomial-time guarantee.
However, this is typical in online algorithms as the focus is given information-theoretic aspects and not on time complexity.
Therefore, an important direction is to 
develop efficient algorithms for the capped objective which could  make the framework more practical and potentially reveal additional structural properties that lead to improved competitive ratios.


\vspace{2mm}
\noindent{\bf Parameter Tuning.} No choice of $\alpha, M,$ and $\theta$ within our interruptible capped-route framework can give $\frac{\mathrm{ALG}}{\mathrm{OPT}}\approx 3.877$ if the checks trigger a cut of the capped route. If no cut under zero-residual instances, we obtained a better ratio. Therefore, improving the ratio $3.877$ for any arbitrary instance needs a new approach which may be radically different from the phase-based approach. 

\vspace{2mm}
\noindent{\bf Improving the Physical-Cost Inequality.} The current analysis uses a worst-case return bound of $u$ after a route prefix of length $u$. On particular instances, the actual endpoint may be much closer to the depot. An endpoint-sensitive launch rule could use this information to reduce the next phase's delay. The main challenge is to charge the resulting delay in the next phase without losing the disjoint tail-allowance accounting. This behavior signals the possibility of running experiments to see how things work in real world situations.

\vspace{2mm}
\noindent{\bf Randomized Interruption Policies.} 
The randomized algorithm were also studied for online $k$-TRP (and $k$-DARP) achieving better expected competitive ratios than the deterministic ones, e.g., \cite{BienkowskiL19-MFCS}. 
In our framework, randomizing the phase offset while retaining the emergency and scheduled interruption rules may improve the expected competitive ratio. The difficulty is that the interruption times and the subsequent cap sequence depend on the random offset. A proof would need to control the expected capped-cost increments and ensure that an allowance from one phase is not used to pay for multiple later interruptions.

\vspace{2mm}
\noindent{\bf Non-Zero Residual Instances.}
We established $(3+\epsilon)$ ratio for zero-residual instances with every $\epsilon>0$. In fact, even when $\sum_j e_j\neq 0$, we can obtain the competitive ratio $(3+\epsilon) \leq (2+2\alpha) (1+\gamma(\alpha)) \leq 3.877$, when $\sum_j e_j\leq \gamma \sum_j\Delta_j \leq \frac{q}{M(q-1)} \mathrm{OPT}$ (Lemma \ref{lemma:ejOPT}). Notice that we cannot set $\alpha=1/2$ to obtain the exact 3 competitive ratio even for when $\sum_je_j=0$ since $\gamma(\alpha)\rightarrow \infty$ because of the factor $(2\alpha-1)$ in the denominator. 
Therefore, it remains open whether exact 3 ratio can be obtained for the zero-residual instances.

\vspace{2mm}
\noindent{\bf Tighter Lower Bounds.} The deterministic upper bound of $3.877$ we established for arbitrary instances here remains separated from the state-of-the-art lower bounds. Establishing stronger adversarial lower bounds, or identifying a matching algorithm, is an important direction. 

\vspace{2mm}
\noindent{\bf Other DARP Variants.} We gave a deterministic $3.877$-guarantee for the uncapacitated $k$-DARP.  For capacitated $k$-DARP, whether preemption is allowed matters. We have provided solutions for the non-preemptive variants with capacity $C\geq 1$ for each server with zero-residual instances.  It remains open whether similar bound can be established for capacitated, preemptive variants.  
Notice that, in preemptive variants, the server(s) may unload a passenger at an intermediate location. However, it changes the capped objective and hence $\Delta_j$ may no longer telescope. Also, it remains open whether the 3.877 ratio can be proven for capacitated DARP for any arbitrary instance.       



\vspace{10mm}
\noindent{\bf AI Statement.} 
The possibility of improving the 4-competitive ratio was suggested in our interactions with ChatGPT -- introduce interruptions. The interruption aspect was later refined to emergency and scheduled cuts, with an emergency cut happening before the scheduled cut. 
Setting $\alpha=0.9$, $M=76$, $\theta=38/5$, and the scheduled cutoff (absolute) time $(1+\alpha)h_j=1.9h_j$, we were able to establish the ratio 3.9 (precisely $2\sqrt{6}-1\approx 3.899$).  We later studied the properties of the algorithm and refined it setting $\alpha\in (1/2,1)$. We then characterized the values of $M$ and $\theta$ w.r.t. $\alpha$ and each other to make the emergency and
scheduled cut cost bounds equal: $M=\frac{\theta}{1-\alpha}$ precisely achieves such cost equality.  This gave us the ratio $\frac{\mathrm{ALG}}{\mathrm{OPT}}$ dependent on $\alpha\in (1/2,1)$. We obtained $3.877$ setting $\alpha\approx 0.892$. 
Our general analysis assuming the cut made us think what if the cut is never triggered. We conducted the modified analysis that established $(3+\epsilon)$ bound with every $\epsilon>0.$
Finally, we used 
AI to improve content and correct grammar in different places. 
The authors take full responsibility for the validity and correctness of the claimed results.

\bibliographystyle{plainurl}
\bibliography{references}

\end{document}